\documentclass[envcountsame]{llncs}

\usepackage{amssymb}
\title{Borel ranks and Wadge degrees of \\ context free  $\om$-languages }
\author{Olivier Finkel\inst{}}
\institute{Equipe de Logique Math\'ematique \\
 U.F.R. de Math\'ematiques, Universit\'e Paris 7 \\ 2 Place Jussieu 75251 Paris
 cedex 05, France \\ \email{finkel@logique.jussieu.fr}.}

\date{}

\begin{document}

\spnewtheorem{The}[theorem]{Theorem}{\bfseries}{\itshape}

\spnewtheorem{Rem}[theorem]{Remark}{\bfseries}{\itshape}
\spnewtheorem{Exa}[theorem]{Example}{\bfseries}{\itshape}
\spnewtheorem{Deff}[theorem]{Definition}{\bfseries}{\itshape}
\spnewtheorem{Lem}[theorem]{Lemma}{\bfseries}{\itshape}
\spnewtheorem{Cor}[theorem]{Corollary}{\bfseries}{\itshape}
\spnewtheorem{Claim}[theorem]{Claim}{\bfseries}{\itshape}
\spnewtheorem{Pro}[theorem]{Proposition}{\bfseries}{\itshape}

\newcommand{\fa}{\forall}
\newcommand{\Ga}{\Gamma}
\newcommand{\Gas}{\Gamma^\star}
\newcommand{\Si}{\Sigma}
\newcommand{\Sis}{\Sigma^\star}
\newcommand{\Sio}{\Sigma^\omega}

\newcommand{\ra}{\rightarrow}
\newcommand{\hs}{\hspace{12mm}

\noi}
\newcommand{\lra}{\leftrightarrow}
\newcommand{\la}{language}
\newcommand{\ite}{\item}
\newcommand{\Lp}{L(\varphi)}
\newcommand{\abs}{\{a, b\}^\star}
\newcommand{\abcs}{\{a, b, c \}^\star}
\newcommand{\ol}{ $\omega$-language}
\newcommand{\orl}{ $\omega$-regular language}
\newcommand{\om}{\omega}
\newcommand{\nl}{\newline}
\newcommand{\noi}{\noindent}
\newcommand{\tla}{\twoheadleftarrow}
\newcommand{\de}{deterministic }
\newcommand{\vp}{\varphi}
\newcommand{\proo}{\noi {\bf Proof.} }
\newcommand {\ep}{\hfill $\square$}

\maketitle

\begin{abstract}
\noi We show that the Borel hierarchy of the class of 
context free $\om$-languages, or even of the class of  $\om$-languages accepted  by B\"uchi $1$-counter automata, 
 is the same as the Borel hierarchy of the class of  $\om$-languages accepted by 
Turing machines with a B\"uchi  acceptance condition. 
In particular, for each recursive non null ordinal $\alpha$,  there exist some 
${\bf \Si}^0_\alpha$-complete and some  ${\bf \Pi}^0_\alpha$-complete 
$\om$-languages accepted by B\"uchi $1$-counter automata.  And the supremum of the set of Borel ranks of 
context free  $\om$-languages is an ordinal $\gamma_2^1$ which is strictly greater than the first non recursive ordinal   
$\om_1^{\mathrm{CK}}$. 
 We then extend this result, proving that the Wadge hierarchy of context free $\om$-languages, 
or even of $\om$-languages accepted  by B\"uchi $1$-counter automata, is the same as the Wadge 
hierarchy of  $\om$-languages accepted  by Turing machines with a B\"uchi or a Muller acceptance condition. 

\end{abstract}

\noi {\small {\bf Keywords:}  $1$-counter B\"uchi automata; 
context free $\omega$-languages; Cantor topology; 
topological properties; Borel hierarchy; Borel ranks; Wadge hierarchy; Wadge degrees.}

\section{Introduction}

\noi Languages of infinite words accepted by finite automata were first studied by B\"uchi 
to prove the decidability of the monadic second order theory of one successor
over the integers. The theory of the so called regular $\om$-languages is now well 
established and has found many applications for specification and verification 
of non-terminating systems; 
see \cite{Thomas90,Staiger97,PerrinPin} for many results and references. 
More powerful  machines, like 
pushdown automata, Turing machines, 
 have also been considered  for the reading of infinite words, 
see Staiger's survey \cite{Staiger97} 
 and the fundamental study \cite{eh}
of Engelfriet and  Hoogeboom on {\bf X}-automata, i.e. finite automata equipped with 
a storage type {\bf X}.  
A way to  study  the complexity of \ol s is to study their topological complexity, 
and particularly to locate them with regard to the Borel and the projective hierarchies.
 On one side  all \ol s  accepted by {\it  \de }
 {\bf X}-automata
with a Muller acceptance condition are  
Boolean combinations of ${\bf \Pi}^0_2$-sets hence 
${\bf \Delta}^0_3$-sets, \cite{Staiger97,eh}.  This implies, from Mc Naughton's Theorem, that 
all regular $\om$-languages, which are accepted by deterministic Muller automata, are also 
${\bf \Delta}^0_3$-sets. 
On the other side, for  {\it  non \de} finite machines, the  
question,  posed by  Lescow and  Thomas in \cite{LescowThomas}, naturally arises: 
what is the topological complexity of \ol s accepted 
by automata equipped with a given storage type {\bf X}? 
 It is well known that every $\om$-language  accepted by a Turing machine (hence also by a 
{\bf X}-automaton) with a Muller acceptance condition is an analytic set. 
In previous papers, we proved that there are context free $\om$-languages, accepted by 
B\"uchi or Muller pushdown automata, 
of every finite Borel rank, of infinite Borel rank, or even being 
analytic but non Borel sets, \cite{DFR,Fin01a,Fin03a,Fin03c}. 
In this paper we show that the Borel hierarchy of $\om$-languages accepted 
by   {\bf X}-automata, for every storage type  {\bf X} such that $1$-counter 
automata can be simulated by {\bf X}-automata, 
 is the same as the Borel hierarchy of $\om$-languages accepted by 
Turing machines with a B\"uchi  acceptance condition. 
In particular, for each recursive non null ordinal $\alpha$,  there exist some 
${\bf \Si}^0_\alpha$-complete and some  ${\bf \Pi}^0_\alpha$-complete 
$\om$-languages accepted by B\"uchi $1$-counter automata, hence also 
in the class ${\bf CFL}_\om$  of   context free $\om$-languages.  
\nl We have to indicate here a mistake in the conference paper  \cite{cie05}. 
We wrote in that paper  that it is well known that if  $L \subseteq \Si^{\om}$   is a (lightface)  $\Si_1^1$ set, i.e. accepted by a Turing machine 
with a B\"uchi acceptance condition,  and is a Borel set  of  rank $\alpha$, then $\alpha$   is smaller than the Church Kleene ordinal $\om_1^{\mathrm{CK}}$, 
which is  the first non recursive ordinal. 
This fact,  which is true if we replace the (lightface)  class  $\Si_1^1$ by the (lightface)  class $\Delta_1^1$,  is actually not true. 
Kechris, Marker and Sami proved in \cite{KMS89} that the supremum 
of the set of Borel ranks of  (lightface) $\Pi_1^1$, so also of  (lightface) $\Si_1^1$,  sets is the ordinal $\gamma_2^1$. 
This ordinal is precisely defined in \cite{KMS89} and it  is strictly greater than the ordinal  $ \om_1^{\mathrm{CK}}$.  
The proofs we give  in   this paper show that  the ordinal $\gamma_2^1$ is also the supremum of the set of Borel ranks of  
$\om$-languages accepted by B\"uchi $1$-counter automata, or of context free $\om$-languages.  
\nl By considering the Wadge hierarchy which is a great refinement of the 
Borel hierarchy, \cite{Wadge83,Duparc01},  we  show the following strengthening of the   preceding result. 
The Wadge hierarchy of the class 
{\bf r}-${\bf BCL}(1)_\om$ of $\om$-languages accepted by real time 
$1$-counter B\"uchi automata, 
 hence also of the class ${\bf CFL}_\om$, 
is the Wadge hierarchy of the class  of $\om$-languages accepted by Turing machines with a B\"uchi acceptance 
condition. 
\nl 
We think that the surprising result obtained in this paper is of interest for both logicians 
working on hierarchies  arising 
in recursion theory or in descriptive set theory, and also for 
computer scientists working on questions connected with non-terminating systems, like the 
construction of effective strategies in infinite games, \cite{Wal,Thomas02,Cachat02,Serre04}. 
\nl The paper is organized as follows. In Section 2 we define multicounter automata which will 
be a useful tool in the sequel. Recall on Borel hierarchy is given in Section 3. 
In Section 4 is studied the Borel hierarchy of $\om$-languages accepted by real time 
$8$-counter automata. The Borel hierarchy of the class {\bf r}-${\bf BCL}(1)_\om$ is studied   in Section 5.  
Results about the Wadge hierarchy of the class {\bf r}-${\bf BCL}(1)_\om$ are given in Section 6.

\section{Multicounter automata}\label{mca}

We assume the reader to be familiar with the theory of formal ($\om$)-languages  
\cite{Thomas90,Staiger97}.
We shall use usual notations of formal language theory. 
\nl  When $\Si$ is a finite alphabet, a {\it non-empty finite word} over $\Si$ is any 
sequence $x=a_1\ldots a_k$, where $a_i\in\Sigma$ 
for $i=1,\ldots ,k$ , and  $k$ is an integer $\geq 1$. The {\it length}
 of $x$ is $k$, denoted by $|x|$.
 The {\it empty word} has no letter and is denoted by $\lambda$; its length is $0$. 
 For $x=a_1\ldots a_k$, we write $x(i)=a_i$  
and $x[i]=x(1)\ldots x(i)$ for $i\leq k$ and $x[0]=\lambda$.
 $\Sis$  is the {\it set of finite words} (including the empty word) over $\Sigma$.
 \nl  The {\it first infinite ordinal} is $\om$.
 An $\om$-{\it word} over $\Si$ is an $\om$ -sequence $a_1 \ldots a_n \ldots$, where for all 
integers $ i\geq 1$, ~
$a_i \in\Sigma$.  When $\sigma$ is an $\om$-word over $\Si$, we write
 $\sigma =\sigma(1)\sigma(2)\ldots \sigma(n) \ldots $,  where for all $i$,~ $\sigma(i)\in \Si$,
and $\sigma[n]=\sigma(1)\sigma(2)\ldots \sigma(n)$  for all $n\geq 1$ and $\sigma[0]=\lambda$.
\nl   The {\it prefix relation} is denoted $\sqsubseteq$: a finite word $u$ is a {\it prefix} 
of a finite word $v$ (respectively,  an infinite word $v$), denoted $u\sqsubseteq v$,  
 if and only if there exists a finite word $w$ 
(respectively,  an infinite word $w$), such that $v=u.w$.
 The {\it set of } $\om$-{\it words} over  the alphabet $\Si$ is denoted by $\Si^\om$.
An  $\om$-{\it language} over an alphabet $\Sigma$ is a subset of  $\Si^\om$.  The complement (in $\Sio$) of an 
$\om$-language $V \subseteq \Sio$ is $\Sio - V$, denoted $V^-$.
 
\begin{Deff} Let $k$ be an integer $\geq 1$. 
A  $k$-counter machine ($k$-CM) is a 4-tuple 
$\mathcal{M}$=$(K,\Si, \Delta, q_0)$,  where $K$ 
is a finite set of states, $\Sigma$ is a finite input alphabet, 
 $q_0\in K$ is the initial state, 
and  $\Delta \subseteq K \times ( \Si \cup \{\lambda\} ) \times \{0, 1\}^k \times K \times \{0, 1, -1\}^k$ is the transition relation. 
The $k$-counter machine $\mathcal{M}$ is said to be {\it real time} iff: 
$\Delta \subseteq K \times
  \Si \times \{0, 1\}^k \times K \times \{0, 1, -1\}^k$, 
 i.e. iff there are not any $\lambda$-transitions. 
\nl  
If  the machine $\mathcal{M}$ is in state $q$ and 
$c_i \in \mathbf{N}$ is the content of the $i^{th}$ counter 
 $\mathcal{C}$$_i$ then 
the  configuration (or global state)
 of $\mathcal{M}$ is the  $(k+1)$-tuple $(q, c_1, \ldots , c_k)$.

\hs For $a\in \Si \cup \{\lambda\}$, 
$q, q' \in K$ and $(c_1, \ldots , c_k) \in \mathbf{N}^k$ such 
that $c_j=0$ for $j\in E \subseteq  \{1, \ldots , k\}$ and $c_j >0$ for 
$j\notin E$, if 
$(q, a, i_1, \ldots , i_k, q', j_1, \ldots , j_k) \in \Delta$ where $i_j=0$ for $j\in E$ 
and $i_j=1$ for $j\notin E$, then we write:
$$a: (q, c_1, \ldots , c_k)\mapsto_{\mathcal{M}} (q', c_1+j_1, \ldots , c_k+j_k)$$

\noi $\mapsto_{\mathcal{M}}^\star$ is the transitive and reflexive closure of
 $\mapsto_{\mathcal{M}}$.
(The subscript $\mathcal{M}$ will be omitted whenever the meaning remains clear).
\nl Thus we see that the transition relation must satisfy:
 \nl if $(q, a, i_1, \ldots , i_k, q', j_1, \ldots , j_k)  \in    \Delta$ and  $i_m=0$ for 
 some $m\in \{1, \ldots , k\}$, then $j_m=0$ or $j_m=1$ (but $j_m$ may not be equal to $-1$).

\hs
Let $\sigma =a_1a_2 \ldots a_n $ be a finite word over $\Si$. 
An sequence of configurations $r=(q_i, c_1^{i}, \ldots c_k^{i})_{1\leq i \leq p}$, for 
$p\geq n+1$, is called 
a  run of $\mathcal{M}$ on $\sigma$, starting in configuration 
$(p, c_1, \ldots, c_k)$, iff:
\begin{enumerate}
\ite[(1)]  $(q_1, c_1^{1}, \ldots c_k^{1})=(p, c_1, \ldots, c_k)$
\ite[(2)] 
 for each $i\geq 1$, there exists $b_i \in \Si \cup \{\lambda\}$ such that 
 $b_i: (q_i, c_1^{i}, \ldots c_k^{i})\mapsto_{\mathcal{M}}
(q_{i+1},  c_1^{i+1}, \ldots c_k^{i+1})$ 
\ite[(3)]  
$a_1.a_2.a_3 \ldots a_n = b_1.b_2.b_3 \ldots b_p$
\end{enumerate}

\noi 
Let $\sigma =a_1a_2 \ldots a_n \ldots $ be an $\om$-word over $\Si$. 
An $\om$-sequence of configurations $r=(q_i, c_1^{i}, \ldots c_k^{i})_{i \geq 1}$ is called 
a run of $\mathcal{M}$ on $\sigma$, starting in configuration 
$(p, c_1, \ldots, c_k)$, iff:
\begin{enumerate}
\ite[(1)]  $(q_1, c_1^{1}, \ldots c_k^{1})=(p, c_1, \ldots, c_k)$

\ite[(2)]   for each $i\geq 1$, there  exists $b_i \in \Si \cup \{\lambda\}$ such that
 $b_i: (q_i, c_1^{i}, \ldots c_k^{i})\mapsto_{\mathcal{M}}  
(q_{i+1},  c_1^{i+1}, \ldots c_k^{i+1})$  
such that either ~  $a_1a_2\ldots a_n\ldots =b_1b_2\ldots b_n\ldots$ 
\nl or ~  $b_1b_2\ldots b_n\ldots$ is a finite prefix of ~ $a_1a_2\ldots a_n\ldots$
\end{enumerate}
\noi The run $r$ is said to be complete when $a_1a_2\ldots a_n\ldots =b_1b_2\ldots b_n\ldots$ 
\nl 
For every such run, $\mathrm{In}(r)$ is the set of all states entered infinitely
 often during run $r$.
\nl
A complete run $r$ of $M$ on $\sigma$, starting in configuration $(q_0, 0, \ldots, 0)$,
 will be simply called ``a run of $M$ on $\sigma$".
\end{Deff}

\begin{Deff} A B\"uchi $k$-counter automaton  is a 5-tuple 
$\mathcal{M}$=$(K,\Si, \Delta, q_0, F)$, 
where $ \mathcal{M}'$=$(K,\Si, \Delta, q_0)$
is a $k$-counter machine and $F \subseteq K$ 
is the set of accepting  states.
The \ol~ accepted by $\mathcal{M}$ is 
\begin{center}
$L(\mathcal{M})$= $\{  \sigma\in\Si^\om \mid \mbox{  there exists a  run r
 of } \mathcal{M} \mbox{ on } \sigma \mbox{  such that } \mathrm{In}(r)
 \cap F \neq \emptyset \}$
\end{center}
\end{Deff}

\begin{Deff} A Muller $k$-counter automaton  is a 5-tuple 
$\mathcal{M}$=$(K,\Si, \Delta, q_0, \mathcal{F})$, 
 where $ \mathcal{M}'$=$(K,\Si, \Delta, q_0)$
is a $k$-counter machine and $\mathcal{F}$$ \subseteq 2^K$ 
is the set of accepting sets of  states.
The \ol~ accepted by $\mathcal{M}$ is 
\begin{center}
$L(\mathcal{M})$=$ \{  \sigma\in\Si^\om \mid \mbox{  there exists a  run r
 of } \mathcal{M}$$ \mbox{ on } \sigma \mbox{  such that } 
\exists F \in \mathcal{F}$$ ~~ \mathrm{In}(r)=F  \}$
\end{center}
\end{Deff}

\noi The class of  B\"uchi $k$-counter automata   will be 
denoted ${\bf BC}(k)$.
\nl The class of real time B\"uchi $k$-counter automata   will be 
denoted {\bf r}-${\bf BC}(k)$.
\nl The class of \ol s accepted by  B\"uchi $k$-counter automata  will be 
denoted ${\bf BCL}(k)_\om$.
\nl The class of \ol s accepted by real time B\"uchi $k$-counter automata  will be 
denoted {\bf r}-${\bf BCL}(k)_\om$.

\hs It is well known that an $\om$-language is accepted by a (real time) 
B\"uchi $k$-counter automaton iff it is accepted by a 
(real time) Muller  $k$-counter automaton \cite{eh}. Notice that it cannot be 
 shown without  using the 
non determinism of automata and this result is no longer true in the  deterministic case.    
\nl  Remark that  $1$-counter automata   introduced above are equivalent to pushdown automata 
whose stack alphabet is in the form $\{Z_0, A\}$ where $Z_0$ is the bottom symbol which always 
remains at the bottom of the stack and appears only there and $A$ is another stack symbol. 
The pushdown stack may be seen like a counter whose content is the integer $N$ if the stack 
content is the word $Z_0.A^N$. 
\nl In the model introduced here the counter value cannot be increased by  more than 1 during  
a single transition. However this does not change the class of $\om$-languages accepted 
by such automata. So the class ${\bf BCL}(1)_\om$ is equal to the class 
{\bf 1}-${\bf ICL_\om}$, introduced in \cite{fin01b},  
and it  is a strict subclass of the class ${\bf CFL}_\om$ of context free \ol s
accepted by B\"uchi pushdown automata.

\section{Borel hierarchy}

\noi We assume the reader to be familiar with basic notions of topology which
may be found in \cite{Moschovakis80,LescowThomas,Kechris94,Staiger97,PerrinPin}.
There is a natural metric on the set $\Sio$ of  infinite words 
over a finite alphabet 
$\Si$ which is called the {\it prefix metric} and defined as follows. For $u, v \in \Sio$ and 
$u\neq v$ let $\delta(u, v)=2^{-l_{\mathrm{pref}(u,v)}}$ where $l_{\mathrm{pref}(u,v)}$ 
 is the first integer $n$
such that the $(n+1)^{st}$ letter of $u$ is different from the $(n+1)^{st}$ letter of $v$. 
This metric induces on $\Sio$ the usual  Cantor topology for which {\it open subsets} of 
$\Sio$ are in the form $W.\Si^\om$, where $W\subseteq \Sis$.
A set $L\subseteq \Si^\om$ is a {\it closed set} iff its complement $\Si^\om - L$ 
is an open set.
Define now the {\it Borel Hierarchy} of subsets of $\Si^\om$:

\begin{Deff}
For a non-null countable ordinal $\alpha$, the classes ${\bf \Si}^0_\alpha$
 and ${\bf \Pi}^0_\alpha$ of the Borel Hierarchy on the topological space $\Si^\om$ 
are defined as follows:
\nl ${\bf \Si}^0_1$ is the class of open subsets of $\Si^\om$, 
 ${\bf \Pi}^0_1$ is the class of closed subsets of $\Si^\om$, 
\nl and for any countable ordinal $\alpha \geq 2$: 
\nl ${\bf \Si}^0_\alpha$ is the class of countable unions of subsets of $\Si^\om$ in 
$\bigcup_{\gamma <\alpha}{\bf \Pi}^0_\gamma$.
 \nl ${\bf \Pi}^0_\alpha$ is the class of countable intersections of subsets of $\Si^\om$ in 
$\bigcup_{\gamma <\alpha}{\bf \Si}^0_\gamma$.
\end{Deff}

\noi For 
a countable ordinal $\alpha$,  a subset of $\Si^\om$ is a Borel set of {\it rank} $\alpha$ iff 
it is in ${\bf \Si}^0_{\alpha}\cup {\bf \Pi}^0_{\alpha}$ but not in 
$\bigcup_{\gamma <\alpha}({\bf \Si}^0_\gamma \cup {\bf \Pi}^0_\gamma)$.

\hs There are also some subsets of $\Si^\om$ which are not Borel.  In particular 
the class of Borel subsets of $\Si^\om$ is strictly included into 
the class  ${\bf \Si}^1_1$ of {\it analytic sets} which are 
obtained by projection of Borel sets, 
see for example \cite{Staiger97,LescowThomas,PerrinPin,Kechris94}
 for more details. 
\nl  We now define completeness with regard to reduction by continuous functions. 
For a countable ordinal  $\alpha\geq 1$, a set $F\subseteq \Si^\om$ is said to be 
a ${\bf \Si}^0_\alpha$  
(respectively,  ${\bf \Pi}^0_\alpha$, ${\bf \Si}^1_1$)-{\it complete set} 
iff for any set $E\subseteq Y^\om$  (with $Y$ a finite alphabet): 
 $E\in {\bf \Si}^0_\alpha$ (respectively,  $E\in {\bf \Pi}^0_\alpha$,  $E\in {\bf \Si}^1_1$) 
iff there exists a continuous function $f: Y^\om \ra \Si^\om$ such that $E = f^{-1}(F)$. 
 ${\bf \Si}^0_n$
 (respectively ${\bf \Pi}^0_n$)-complete sets, with $n$ an integer $\geq 1$, 
 are thoroughly characterized in \cite{Staiger86a}.  

\noi The (lightface) class $\Si^1_1$ of {\it effective analytic sets} 
is the class of sets which are obtained by projection of arithmetical sets. It is 
well known that a set $L \subseteq \Sio$, where $\Si$ is a finite alphabet, 
 is in the class $\Si^1_1$  iff it is accepted by a Turing machine with a B\"uchi or Muller 
acceptance condition \cite{Staiger97}. 
\nl  As indicated in the introduction, we made a  mistake in the conference paper  \cite{cie05}. 
We wrote there  that it is well known that if  $L \subseteq \Si^{\om}$   is a (lightface)  $\Si_1^1$ set, 
and is a Borel set  of  rank $\alpha$, then $\alpha$   is smaller than $\om_1^{\mathrm{CK}}$. 
This fact,  which is true if we replace the (lightface)  class  $\Si_1^1$ by the (lightface)  class $\Delta_1^1$,  is actually not true. 
Kechris, Marker and Sami proved in \cite{KMS89} that the supremum 
of the set of Borel ranks of  (lightface) $\Pi_1^1$, so also of  (lightface) $\Si_1^1$,  sets is the ordinal $\gamma_2^1$. 
\nl This ordinal is precisely defined in \cite{KMS89}.  Kechris, Marker and Sami proved that the ordinal $\gamma_2^1$
 is strictly greater than the ordinal $\delta_2^1$ which is the first non $\Delta_2^1$ ordinal. 
Thus in particular it holds that $ \om_1^{\mathrm{CK}} < \gamma_2^1$.  
The exact value of the ordinal $\gamma_2^1$ may depend on axioms of  set theory  \cite{KMS89}. It is consistent with the axiomatic system 
{\bf ZFC } that $\gamma_2^1$ is equal to the ordinal $\delta_3^1$ which is the first non $\Delta_3^1$ ordinal  
(because $\gamma_2^1= \delta_3^1$  in {\bf ZFC + (V=L)}). On the other hand  the axiom of 
$\Pi_1^1$-determinacy implies that $\gamma_2^1 < \delta_3^1$.  For more details,  the reader is  referred to  \cite{KMS89} and to 
a textbook of set theory like \cite{Jech}. 

\hs Notice however that it seems still unknown  whether {\it every } non null ordinal $\gamma < \gamma_2^1$ is the Borel rank 
of a (lightface) $\Pi_1^1$ (or $\Si_1^1$) set. 
On the other hand it is known that every ordinal $\gamma < \om_1^{\mathrm{CK}}$ is the Borel rank 
of a (lightface) $\Delta_1^1$ set.  Moreover, for every non null  ordinal $\alpha < \om_1^{\mathrm{CK}}$, there exist some  
 ${\bf \Si}^0_\alpha$-complete and some  ${\bf \Pi}^0_\alpha$-complete sets in the class $\Delta_1^1$. 
Louveau gives the following argument: the natural universal set for the class ${\bf \Si}^0_\alpha$ (respectively,  ${\bf \Pi}^0_\alpha$), where 
$\alpha <  \om_1^{\mathrm{CK}}$, 
is a  ${\bf \Si}^0_\alpha$-complete (respectively, ${\bf \Pi}^0_\alpha$-complete) set and it is in the class $\Delta_1^1$, \cite{Louveau05}. 
The definition and the construction of a universal set for a given Borel class may be found in \cite{Moschovakis80}.

\section{Borel hierarchy of  $\om$-languages in {\bf r}-${\bf BCL}(8)_\om$}\label{section4}

\noi It is well known that every Turing machine can be simulated by a 
(non real time) $2$-counter automaton, see \cite{HopcroftUllman79}.  
Thus the Borel hierarchy of the class  ${\bf BCL}(2)_\om$ is also the Borel hierarchy of the class 
of $\om$-languages accepted by B\"uchi Turing machines. 
We shall  prove the following proposition. 

\begin{Pro} 
The Borel hierarchy of the class  {\bf r}-${\bf BCL}(8)_\om$ is equal to the Borel hierarchy of the class ${\bf BCL}(2)_\om$. 
\end{Pro}

\noi  We first sketch the proof of this result.  We are going to find, from an $\om$-language $L \subseteq \Sio$ in ${\bf BCL}(2)_\om$, another 
$\om$-language $\theta_S(L)$ which will be of the same Borel complexity but  accepted by a {\it real-time } 8-counter B\"uchi automaton.  
The idea is to add firstly a storage type called a queue to a 2-counter B\"uchi automaton in order to read $\om$-words in real-time. Then we shall 
see that a queue can be simulated by two pushdown stacks or by four counters. This simulation is not done in real-time but a crucial fact is that we can bound 
the number of transitions needed to simulate the queue. This allows to pad the strings in $L$ with enough extra letters so that the new words will be read in 
real-time by a 8-counter B\"uchi automaton (two  counters are used to check that an $\om$-word is really obtained with the good padding 
which is made in a regular way).  The padding is obtained via the function $\theta_S$ which we define now. 

\hs Let $\Si$ be an alphabet having at least two letters,  $E$ be a new letter not in 
$\Si$,  $S$ be an integer $\geq 1$, and $\theta_S: \Sio \ra (\Sigma \cup \{E\})^\om$ be the 
function defined, for all  $x \in \Sio$, by: 
$$ \theta_S(x)=x(1).E^{S}.x(2).E^{S^2}.x(3).E^{S^3}.x(4) \ldots 
x(n).E^{S^n}.x(n+1).E^{S^{n+1}} \ldots $$

\hs We now  state  the two  following lemmas. 

\begin{Lem}\label{thetaborel}
Let $\Si$ be an alphabet having at least two letters and let $L \subseteq \Sio$ be a subset 
of $\Sio$ which is ${\bf \Si}^0_\alpha$-complete (respectively,   ${\bf \Pi}^0_\alpha$-complete,     ${\bf \Si}^0_\alpha$ of rank $\alpha$,    
 ${\bf \Pi}^0_\alpha$  of rank $\alpha$) for some ordinal $\alpha \geq 2$. Then  the 
$\om$-language $\theta_S(L)$ is a  subset of $(\Sigma \cup \{E\})^\om$ which is ${\bf \Si}^0_\alpha$-complete 
(respectively,   ${\bf \Pi}^0_\alpha$-complete,     ${\bf \Si}^0_\alpha$ of rank $\alpha$,    
 ${\bf \Pi}^0_\alpha$  of rank $\alpha$). 
\end{Lem}

\begin{proof}
 Let $\Si$ be an alphabet having at least two letters.  
 It is easy to see that the function  $\theta_S$ is continuous because if 
two $\om$-words $x$ and $y$ of $\Sio$ have a common initial segment of length $n$ then 
the two $\om$-words $\theta_S(x)$ and $\theta_S(y)$
have a common initial segment of length greater than $n$. 
\nl Let $\vp : (\Sigma \cup \{E\})^\om \ra (\Sigma \cup \{E\})^\om$ be the mapping 
defined, for all $y\in (\Sigma \cup \{E\})^\om$, by: 
$\vp(y)=y(1).y(S+2).y(S+S^2+3)\ldots y(S+S^2+\ldots +S^n+(n+1))\ldots$ 
\nl It is easy to see that the function $\vp$ is also continuous and that, for any 
$L \subseteq \Sio$, 
$\theta_S(L)=\vp^{-1}(L) \cap \theta_S(\Sio)$. 
\nl Let  now $L \subseteq \Sio$ be a 
${\bf \Si}^0_\alpha$  (respectively,   ${\bf \Pi}^0_\alpha$) subset 
of $\Sio$, hence also of  $(\Sigma \cup \{E\})^\om$, 
  for some ordinal $\alpha \geq 2$. Then $\vp^{-1}(L)$ is a 
${\bf \Si}^0_\alpha$ (respectively,  ${\bf \Pi}^0_\alpha$) subset of 
$(\Sigma \cup \{E\})^\om$ because the class ${\bf \Si}^0_\alpha$ 
(respectively,  ${\bf \Pi}^0_\alpha$) is closed under inverse images by continuous 
functions.  On the other hand $\theta_S(\Sio)$ is a closed set thus 
$\theta_S(L)=\vp^{-1}(L) \cap \theta_S(\Sio)$ is  a 
${\bf \Si}^0_\alpha$ (respectively,  ${\bf \Pi}^0_\alpha$) subset of 
$(\Sigma \cup \{E\})^\om$ because the class  ${\bf \Si}^0_\alpha$ 
(respectively,  ${\bf \Pi}^0_\alpha$) is closed under finite intersection. 

\noi Moreover it holds that $L=\theta_S^{-1}(\theta_S(L))$. Thus if 
$L$ is  assumed to be ${\bf \Si}^0_\alpha$-complete (respectively,   ${\bf \Pi}^0_\alpha$-complete) then 
 $\theta_S(L)$ is  a 
${\bf \Si}^0_\alpha$-{\bf complete} (respectively,  
${\bf \Pi}^0_\alpha$-{\bf complete} ) subset of 
$(\Sigma \cup \{E\})^\om$ because it is a  ${\bf \Si}^0_\alpha$ 
(respectively,  ${\bf \Pi}^0_\alpha$) set, the function $\theta_S$ is continuous, and 
$L$ is  ${\bf \Si}^0_\alpha$-complete (respectively,   ${\bf \Pi}^0_\alpha$-complete). 

\noi Assume now that $L$ is a ${\bf \Si}^0_\alpha$-set of rank $\alpha$,    
(respectively,   ${\bf \Pi}^0_\alpha$-set  of rank $\alpha$). We have already seen that 
$\theta_S(L)$ is  a 
${\bf \Si}^0_\alpha$ (respectively,  ${\bf \Pi}^0_\alpha$) subset of 
$(\Sigma \cup \{E\})^\om$. The Borel rank of $\theta_S(L)$  cannot be smaller than  $\alpha$ because otherwise 
$L=\theta_S^{-1}(\theta_S(L))$ would be also of Borel rank smaller than  $\alpha$.

\ep  \end{proof}

\begin{Lem}\label{theta_S}
Let $\Si$ be an alphabet having at least two letters and let $L \subseteq \Sio$ be an 
$\om$-language in the class  ${\bf BCL}(2)_\om$. Then there exists an integer $S \geq 1$ such 
that $\theta_S(L)$ is in the class  {\bf r}-${\bf BCL}(8)_\om$. 
\end{Lem}

\begin{proof}
Let $\Si$ be an alphabet having at least two letters and let $L \subseteq \Sio$ be an 
$\om$-language accepted by  a B\"uchi $2$-counter automaton $\mathcal{A}$. 

\hs A way to construct a finite machine accepting the same $\om$-language but 
 being {\it real time} would be  to add a storage type called a {\it queue} \cite{eh}. 

\hs  Configurations 
of a queue are finite words over a finite alphabet $\Si$; 
a letter of $\Si$ may be added to the rear of the queue or removed 
from the front; moreover there are tests to determine the first letter of the queue. 

\hs The new machine will read words in  real time. At every transition a letter of the input 
$\om$-word is read, is added to the rear of  the queue, waiting to be read 
(and then removed from the front of the queue)  for the simulation of 
the reading of the input $\om$-word by the $2$-counter automaton $\mathcal{A}$.

\hs We are going to see that one can simulate a queue with four counters. This simulation 
will not be a {\it real time} simulation but we shall be able to get an upper bound 
on  the number 
of transitions of the four counters which are necessary for the simulation of one 
transition of the 
queue. This upper bound will be useful in the sequel for our purpose. 

\begin{Claim}\label{C1}
A queue can be simulated by two pushdown stacks. 
\end{Claim}

\begin{proof} Assume that the queue alphabet is 
$\Si=\{Z_2, Z_3, \ldots , Z_{k-1} \}$, for some integer $k\geq 3$.  
\nl The content of the queue can be represented by a finite word 
$Z_{i_1}Z_{i_2}Z_{i_3}    \ldots   Z_{i_m}$, 
 where the letter $Z_{i_m}$ is the first letter of the queue and $Z_{i_1}$ is the 
 last letter of the queue (the last added  to the rear). 
\nl This content can be stored in a pushdown stack whose alphabet is 
$\Ga=\Si\cup\{Z_1\}$, where $Z_1$ is the bottom symbol which appears only at the bottom of 
the stack and always remains there. The stack content representing the queue content 
$Z_{i_1}Z_{i_2}Z_{i_3}    \ldots   Z_{i_m}$ will be simply 
$Z_1 Z_{i_1} Z_{i_2} Z_{i_3}    \ldots   Z_{i_m}$, where 
$Z_1$ is at the bottom of the stack and $Z_{i_m}$ is at the top of the stack. 
\nl If the letter $Z_{i_m}$ of the front of the queue is removed from the queue, it suffices 
to pop  the same letter from  the top of the stack. 
\nl To simulate the addition of a new letter $Z_r$ to the rear of the queue we can use a 
second pushdown stack whose alphabet is also $\Ga$. 
\nl  we have in fact  to add the letter $Z_r$
between the letters $Z_1$ and $Z_{i_1}$ of the first  pushdown stack. 
To achieve this, we successively pop letters from the top 
of the first stack, pushing them in the second stack containing only 
at the beginning the bottom 
symbol $Z_1$. After having done this operation for letters 
$Z_{i_m}, Z_{i_{m-1}}, \ldots Z_{i_2}, Z_{i_1}$, the content of the first stack is 
$Z_1$ and the content of the second stack is $Z_1 Z_{i_m}Z_{i_{m-1}}\ldots Z_{i_2}Z_{i_1}$. 
We can then push the letter $Z_r$ at the top of the first stack. Then we successively 
pop letters $Z_{i_1},  Z_{i_2},  Z_{i_3}    \ldots  ,  Z_{i_m}$ from the top 
of the second stack, pushing them in the first stack. 
At the end of this operation the content of the first stack is 
$Z_0Z_r Z_{i_1} Z_{i_2} Z_{i_3}    \ldots   Z_{i_m}$ and it represents 
the new content of the queue. 
\ep  \end{proof}

\hs We recall now the following well known property. 

\begin{Claim}[\cite{HopcroftUllman79}]\label{C4}
A pushdown stack can 
be simulated by two counters.   
\end{Claim}

\begin{proof} Consider a stack having $k-1$ symbols $Z_1, Z_2, \ldots , Z_{k-1}$.  The stack 
content $Z_{i_1} Z_{i_2} Z_{i_3}    \ldots   Z_{i_m}$ can be represented by the integer $j$
which is given  in base $k$ by:
$$j = i_m + k.i_{m-1} + k^2.i_{m-2} + \ldots + k^{m-1}.i_1$$
\noi Notice that, as remarked in  \cite{HopcroftUllman79}, 
not every integer represents a stack content. 
In particular an integer whose representation in base $k$ contains the digit $0$ does not 
represent any stack content.

\hs We are going to see how to use a second counter to determine which is the letter 
at the top of the stack, and to simulate the operations of 
pushing a letter in the stack or of popping a letter from the top of the stack. 

\hs Assume that the integer $j$ representing the stack content 
$Z_{i_1} Z_{i_2} Z_{i_3}    \ldots   Z_{i_m}$ is stored in one of the two counters. 

\hs In order to determine which is the letter at the top of the stack, we can copy 
the content $j$ in the second counter, using the finite control of the finite machine 
to compute $j$ modulo $k$. 
\nl The integer  $j$ modulo $k$ is equal to the integer $i_m$ which 
characterizes the letter $Z_{i_m}$ hence the letter at the top of the stack. 
\nl It would be possible to transfer again the integer $j$ in the first counter, but 
one can also leave it in the second counter and use the finite control to know
in which counter is stored the integer $j$. 
\nl Notice that this operation needs only $j$ steps (and $2j$ steps if we transfer 
again $j$ in the first counter). 

\hs If a letter $Z_r$ is pushed into the stack, the content of the stack is now
 $Z_{i_1} Z_{i_2} Z_{i_3}    \ldots   Z_{i_m}Z_r$ and the integer associated with that 
content is $j.k + r$. 
\nl It is easy to store the integer $j.k$ in the second counter, by adding 
$k$ to this counter each time  the first counter is decreased by $1$. 
When the content of the first counter is equal to zero then the content of the second counter 
is equal to $j.k$. One can then add $r$ to the second counter by using the finite control 
of the machine. Again we can use the finite control to know that now the  content of the stack 
is coded by the integer which is  in the {\it second } counter. 
\nl Notice that the whole operation needs only $j.k + r$ steps. 

\hs  If instead the symbol $Z_{i_m}$ is popped from the top of the stack, then the new content 
of the stack is $Z_{i_1} Z_{i_2} Z_{i_3}    \ldots   Z_{i_{m-1}}$ and it is represented by 
the integer $[\frac{j}{k}] = i_{m-1} + k.i_{m-2} + \ldots + k^{m-2}.i_1$ 
which is the integer part of $\frac{j}{k}$.  
\nl To get the integer $[\frac{j}{k}]$ as content of the second counter, 
we can decrease the first 
counter from $j$ to zero, adding $1$ to the second counter each time the first one 
is decreased by $k$. 
\nl Notice that this  operation needs only $j$ steps. 
\nl Remark also that we can achieve this operation in a non deterministic way, checking 
at the end of it which was the letter at the top of the stack. 
\ep  \end{proof} 

\hs  We have seen above 
that a queue can be simulated with two pushdown stacks hence also  with four counters. 
\nl  This simulation is not done in {\it real time} but we shall see that we can 
get an upper bound on the number of transitions of the four 
counters simulating one transition of the queue.  This upper bound will be crucial in view of 
Lemma \ref{theta_S}.

\begin{Claim}\label{C5}
Assume as above that the queue alphabet is 
$\Si=\{Z_2, Z_3, \ldots , Z_{k-1} \}$, for some integer $k\geq 3$ and that at some time 
the content of the queue is a finite word 
$Z_{i_1}Z_{i_2}Z_{i_3}    \ldots   Z_{i_m}$. 
Then the number of transitions of  four counters which are needed 
to simulate the addition of a  letter $Z_r$ to the rear of the queue is smaller 
than $(2.k)^{m+2}$. 
\end{Claim}

\begin{proof}
Recall that the content of the queue  can be stored in a pushdown stack whose alphabet is 
$\Ga=\Si\cup\{Z_1\}$, where $Z_1$ is the bottom symbol. 
The stack content representing the queue content 
$Z_{i_1}Z_{i_2}Z_{i_3}    \ldots   Z_{i_m}$ is  simply 
$Z_1 Z_{i_1} Z_{i_2} Z_{i_3}    \ldots   Z_{i_m}$, where 
$Z_1$ is at the bottom of the stack and $Z_{i_m}$ is at the top of the stack. 
\nl This stack 
content $Z_1 Z_{i_1} Z_{i_2} Z_{i_3}    \ldots   Z_{i_m}$ can itself 
be represented by the integer $j$
which is given  in base $k$ by:
$$j = i_m + k.i_{m-1} + k^2.i_{m-2} + \ldots + k^{m-1}.i_1 + k^m.1 ~ \leq ~k^{m+1}$$

\noi We have seen that, considering the simulation of the addition of $Z_r$ to the rear of 
a queue with two pushdown stacks, the first 
stack containing  $Z_1 Z_{i_1} Z_{i_2} Z_{i_3} \ldots Z_{i_{m}}$, we have first 
to successively pop letters $Z_{i_{m}}, \ldots,  Z_{i_3}, Z_{i_2}, Z_{i_1}$,  
from the top of the first stack 
and push them in the second stack. 
\nl  We have also seen above that  popping a letter from 
the stack whose content is represented by the integer $j$ needs only $j$ transitions 
of two counters and that we can know at the end of this popping simulation which letter 
has just been popped. 
\nl Moreover to push a letter $Z_s$ in the  stack whose 
content is represented by an integer $j'$ needs only $j'.k + s$ 
transitions of two counters. 
\nl Thus at most  $2.m.k^{m+1}$ transitions of  four  counters are necessary  to simulate 
the operation of successively 
popping  letters $Z_{i_{m}}, \ldots,  Z_{i_3}, Z_{i_2}, Z_{i_1}$,  
from the top of the first stack 
and then pushing  them in the second stack. 
\nl Two transitions of the counters are needed to check that the content of the first stack is now 
reduced to $Z_1$, which is simply represented by the integer $1$, without changing this 
content; in one step the counter 
is reduced from one to zero then in a second step the counter is increased from zero to one. 
\nl To simulate the addition of  the letter $Z_r$ to the rear of the queue, we now 
push the letter $Z_r$ in the first stack; this is simulated by $k + r$ 
transitions of two counters. 
\nl Now we have to successively 
pop letters $Z_{i_{1}}, Z_{i_2}, Z_{i_3}, \ldots,  Z_{i_m}$, 
 from the top of the second  stack 
and push them again in the first stack. 
This whole operation needs less than 
 $m.k^{m+1} + m.k^{m+2}$ transitions of the  four  counters. 

\hs Finally, to simulate   the addition of a  letter $Z_r$ to the rear of the queue, we need 
only   
$$2.m.k^{m+1} +  2 + k + r +  m.k^{m+1} + m.k^{m+2}$$ 
\noi transitions of four counters.  This number is smaller than 
$$4.m.k^{m+2} + 3.k  \leq (2.k)^{m+2}.$$ 
\ep  \end{proof} 

\begin{Claim}\label{C6}
Assume as above that the queue alphabet is 
$\Si=\{Z_2, Z_3, \ldots , Z_{k-1} \}$, for some integer $k\geq 3$ and that at some time 
the content of the queue is a finite word 
$Z_{i_1}Z_{i_2}Z_{i_3}    \ldots   Z_{i_m}$. 
Then the number of transitions of  four counters which are needed 
to determine the letter $Z_{i_m}$ which is at the front  of the queue is smaller   
than $k^{m+1}$. And the number of transitions of  four counters which are needed 
to simulate the operation of removing 
 the letter $Z_{i_m}$ from the front  of the queue is smaller than  $k^{m+1}$. 

\end{Claim}

\begin{proof} The content $Z_{i_1}Z_{i_2}Z_{i_3}    \ldots   Z_{i_m}$ of the queue  is represented by 
a stack content 
$Z_1 Z_{i_1} Z_{i_2} Z_{i_3}    \ldots   Z_{i_m}$, where 
$Z_1$ is at the bottom of the stack and $Z_{i_m}$ is at the top of the stack. 
\nl This stack 
content is itself represented by the integer 
$$j = i_m + k.i_{m-1} + k^2.i_{m-2} + \ldots + k^{m-1}.i_1 + k^m.1 ~ \leq ~k^{m+1}$$
\noi By the proof of Claim \ref{C4}, $j \leq k^{m+1}$ transitions of four counters 
(and even of only two counters) 
suffice to determine the letter $Z_{i_m}$ which is at the top of the stack or to pop
it from the top of the stack. 
\ep  \end{proof} 

\begin{Claim}\label{C7}
The $\om$-language $\theta_S(\Sio)$ is in the class {\bf r}-${\bf BCL}(2)_\om$. 
\end{Claim}

\begin{proof} Recall that if 
 $\Si$ is an alphabet having at least two letters,  $E$ is a new letter not in 
$\Si$,  $S$ is an integer $\geq 1$, then an $\om$-word $y \in (\Sigma \cup \{E\})^\om$ is in 
$\theta_S(\Sio)$ iff it is in the form: 
$$ \theta_S(x)=x(1).E^{S}.x(2).E^{S^2}.x(3).E^{S^3}.x(4) \ldots 
x(n).E^{S^n}.x(n+1).E^{S^{n+1}} \ldots $$
\noi for some $x \in \Sio$. 
\nl It is easy to construct a real time B\"uchi $2$-counter automaton 
$\mathcal{B}$ accepting $\theta_S(\Sio)$. 
We describe now the behaviour of $\mathcal{B}$ when reading 
an $\om$-word $y \in (\Sigma \cup \{E\})^\om$.  
After the reading of the first letter $y(1)\in \Si$, the automaton $\mathcal{B}$ adds one 
to the first counter for each letter $E$ read, checking with the finite control that there 
are $S$ letters $E$ following $y(1)$. Then $\mathcal{B}$ reads a second letter of $\Si$ and 
next it adds $S$ to the second counter and decreases the first counter by one each time it 
reads $S$ letters $E$. When the first counter content is equal to zero, the second counter 
content is equal to  $S^2$ and $\mathcal{B}$ has read 
$S^2$ letters $E$. It then reads a third letter of $\Si$, and next 
it adds $S$ to the first counter and decreases the second  counter by one each time it 
reads $S$ letters $E$. When the second  counter content is equal to zero, the first counter 
content is equal to $S^3$ and 
$\mathcal{B}$ has read $S^3$ letters $E$. Then $\mathcal{B}$  reads a fourth letter of $\Si$, 
and so on. The B\"uchi acceptance condition is used to check that the content of the first 
(and also of the second) counter takes infinitely many times the value zero. 
\ep  \end{proof}

 {\it End of  Proof of Lemma \ref{theta_S}}.  
Let $L \subseteq \Sio$ be an $\om$-language accepted by a B\"uchi $2$-counter automaton 
$\mathcal{A}$. We are going to explain the behaviour of a real time 
B\"uchi $8$-counter automaton $\mathcal{A}$$_1$ accepting $\theta_S(L)$ where  $S=(3k)^3$ 
and $k=cardinal(\Si)+2$. 
\nl  As explained in the proof of preceding Claim \ref{C7}, two counters of $\mathcal{A}$$_1$
will be  used, independently of the other six ones,  
to check that the input  $\om$-word $y \in (\Sigma \cup \{E\})^\om$ is in 
$\theta_S(\Sio)$. 
\nl  Consider now the reading by $\mathcal{A}$$_1$ of an $\om$-word 
$y \in (\Sigma \cup \{E\})^\om$ in the form $y = \theta_S(x) = 
x(1).E^{S}.x(2).E^{S^2}.x(3).E^{S^3}.x(4) \ldots 
x(n).E^{S^n}.x(n+1).E^{S^{n+1}} \ldots $ for some $x\in \Sio$. 
\nl  The automaton $\mathcal{A}$$_1$ will simulate, using four counters, 
 a queue in which will be successively stored letters 
$x(1), x(2), \ldots, x(n), \ldots$ as soon as they will be read. Two other counters 
of $\mathcal{A}$$_1$ will be used to simulate the reading of the $\om$-word $x$ by the 
B\"uchi $2$-counter automaton $\mathcal{A}$. 
 Notice that only letters $x(1), x(2), \ldots, x(n), \ldots$ will be added to the rear 
of the queue. Therefore after having read the initial segment 
 $x(1).E^{S}.x(2).E^{S^2}.x(3).E^{S^3}.x(4) \ldots 
x(n).E^{S^n}$ 
 of $y$ the content of the queue has cardinal smaller than or equal to 
$n$. 
\nl When $\mathcal{A}$$_1$ reads $x(n+1)$ it will firstly simulate the addition of the letter 
$x(n+1)$ to the rear of the queue, using four counters, and doing this in real time while 
continuing to read some following letters $E$. 
By Claim \ref{C5}, the number of transitions of four counters needed to simulate 
the addition of $x(n+1)$ to the rear of the queue is smaller  than $(2k)^{n+2}$ where 
$k=cardinal(\Si)+2$. 
Next $\mathcal{A}$$_1$ determines which is the letter at the front of the queue. By Claim 
\ref{C6} this needs at most $k^{n+2}$ transitions of the four counters (because there 
are now at most $(n+1)$ letters in the queue). Now the automaton 
 $\mathcal{A}$$_1$ simulates, using two counters,  only one transition of $\mathcal{A}$. 
This transition  may be a $\lambda$-transition or not. In the second case $\mathcal{A}$$_1$ 
simulates the reading by $\mathcal{A}$ of the letter 
at the front of the queue so this letter 
is removed from the queue; by Claim \ref{C6} this needs again  at most 
$k^{n+2}$ transitions of the four counters. 
 It holds that 
$$(2k)^{n+2} + k^{n+2}  + k^{n+2} + 1 \leq ((3k)^3)^{n}$$
\noi Thus if we set $S=(3k)^3$ then all these transitions of the six counters can be achieved
by the automaton $\mathcal{A}$$_1$ in {\it real-time} during the reading of the letters 
$E$ following $x(n+1)$ in $y$. Not all the $S^{n+1}$ letters $E$ are read during 
these transitions of the six counters. But $\mathcal{A}$$_1$ will read the other ones without 
changing the contents of the six counters, waiting for the reading of the next letter of 
$\Si$: the letter $x(n+2)$. It will then simulate the addition of this letter to the rear 
of the queue, and so on. 
\nl A  Muller condition can be used to ensure that $y \in \theta_S(\Sio)$, i.e. 
$y=\theta_S(x)$ for some $x\in \Sio$, {\it and}  that $x\in L=L(\mathcal{A})$. 
As mentioned in Section \ref{mca}, this can also be achieved  
with a B\"uchi acceptance condition. 
\ep  \end{proof}

\hs Notice that we did not aim to find the smallest possible  integer $S$ but only to find 
one such integer in order to prove Lemma \ref{theta_S}.

\section{Borel hierarchy of  $\om$-languages in {\bf r}-${\bf BCL}(1)_\om$}

\noi  We shall firstly prove the following result. 

\begin{Pro}\label{prop} Let $k\geq 2$ be an integer. 
If,  for some ordinal $\alpha \geq 2$,  
there is an $\om$-language in the class  {\bf r}-${\bf BCL}(k)_\om$ which is 
 ${\bf \Si}^0_\alpha$-complete (respectively,   ${\bf \Pi}^0_\alpha$-complete,  ${\bf \Si}^0_\alpha$ of rank $\alpha$,  
 ${\bf \Pi}^0_\alpha$ of rank $\alpha$),  then 
there is some $\om$-language in the class  {\bf r}-${\bf BCL}(1)_\om$ which is 
 ${\bf \Si}^0_\alpha$-complete (respectively,   ${\bf \Pi}^0_\alpha$-complete, ${\bf \Si}^0_\alpha$ of rank $\alpha$,  
 ${\bf \Pi}^0_\alpha$ of rank $\alpha$). 
\end{Pro}

\noi To simplify the exposition of the proof of this result, firstly, we are going  to give 
the proof for $k=2$. Next we shall explain the modifications to do in order to infer 
the result for the integer $k=8$ which is in fact the only case we shall need in the sequel. 
(However our main result will show that the proposition is true for every integer $k\geq 2$).
\nl
 For that purpose we define first a coding of $\om$-words 
over a finite alphabet $\Si$ by $\om$-words over the alphabet  $\Si\cup\{A, B, 0\}$ where  
$A$, $B$ and $0$  are  new letters not in $\Si$.
We shall code an $\om$-word $x\in \Si^{\om}$ by the $\om$-word $h(x)$ defined by
$$h(x)=A.0^6.x(1).B.0^{6^2}.A.0^{6^2}.x(2).B.0^{6^3}.A.0^{6^3}.x(3).B \ldots  
B.0^{6^n}.A.0^{6^n}.x(n).B \ldots  $$

\noi   This coding defines a  mapping  $h: \Si^{\om} \ra (\Si\cup\{A, B, 0\})^\om$. The 
function $h$ is continuous because for all $\om$-words $x, y \in \Si^{\om}$ and 
each positive integer $n$,  it holds 
that  $\delta(x, y) < 2^{-n} \ra \delta( h(x), h(y) ) < 2^{-n}$.

\begin{Lem}\label{lem3}
Let $\Si$ be a finite alphabet and 
$(h(\Si^{\om}))^- =  (\Si \cup\{A, B, 0\})^\om - h(\Si^{\om})$. 
If $\mathcal{L}$$ \subseteq \Si^{\om}$ is  
${\bf \Si}^0_\alpha$-complete (respectively, ${\bf \Pi}^0_\alpha$-complete, ${\bf \Si}^0_\alpha$ of rank $\alpha$,  
 ${\bf \Pi}^0_\alpha$ of rank $\alpha$), for a countable 
ordinal $\alpha \geq 2$, then 
$h(\mathcal{L}) $$\cup h(\Si^{\om})^-$
 is a subset of $(\Si \cup\{A, B, 0\})^\om$ which is 
${\bf \Si}^0_\alpha$-complete (respectively, ${\bf \Pi}^0_\alpha$-complete, ${\bf \Si}^0_\alpha$ of rank $\alpha$,  
 ${\bf \Pi}^0_\alpha$ of rank $\alpha$).
\end{Lem}

\begin{proof}
  The topological space $\Si^{\om}$  is compact 
thus its image by the continuous function 
$h$ is also a compact subset of the topological space 
$(\Si \cup\{A, B, 0\})^\om$. 
The set  $h(\Si^{\om})$ is compact hence  it is a closed subset of 
$(\Si \cup\{A, B, 0\})^\om$. Then its complement 
$$(h(\Si^{\om}))^- =  (\Si \cup\{A, B, 0\})^\om - h(\Si^{\om})$$
\noi  is an open (i.e. a ${\bf \Si^0_1}$) subset of $(\Si \cup\{A, B, 0\})^\om$.

 \hs   On the other hand the function $h$ is also injective 
thus it is a bijection from $\Si^{\om}$  onto 
$h(\Si^{\om})$. But a continuous bijection between two compact sets is an homeomorphism
therefore $h$ induces an homeomorphism between  $\Si^{\om}$  and  $h(\Si^{\om})$. 
Assume that  $\mathcal{L}$ is a ${\bf \Si}^0_\alpha$
(respectively, ${\bf \Pi}^0_\alpha$) subset 
of $\Si^{\om}$. Then    
$h(\mathcal{L})$ is a 
${\bf \Si}^0_\alpha$ (respectively, ${\bf \Pi}^0_\alpha$)  subset of  $h(\Si^{\om})$ 
(where Borel sets of the topological 
space $h(\Si^{\om})$ are defined from open sets as in the case of the topological 
space $\Sio$). 

\hs    The topological space $h(\Si^{\om})$ is a 
topological subspace of $ (\Si \cup\{A, B, 0\})^\om$ and its 
topology  is induced by the topology on $(\Si \cup\{A, B, 0\})^\om$: open sets 
of $h(\Si^{\om})$ are traces on $h(\Si^{\om})$ of open sets of 
$(\Si \cup\{A, B, 0\})^\om$ and the same result holds for closed sets. Then 
one can easily show by induction that for every ordinal  $\alpha \geq 1$,  
${\bf \Pi}^0_\alpha $-subsets 
(resp.  ${\bf \Si}^0_\alpha $-subsets) of  $h(\Si^{\om})$ are traces on $h(\Si^{\om})$ of 
${\bf \Pi}^0_\alpha $-subsets 
(resp.  ${\bf \Si}^0_\alpha $-subsets) of  $(\Si \cup\{A, B, 0\})^\om$, i.e. are 
intersections with $h(\Si^{\om})$ of 
${\bf \Pi}^0_\alpha $-subsets 
(resp.  ${\bf \Si}^0_\alpha $-subsets) of  $(\Si \cup\{A, B, 0\})^\om$. 

\hs    But  $h(\mathcal{L})$ is a ${\bf \Si}^0_\alpha$ 
(respectively, ${\bf \Pi}^0_\alpha$)-subset 
of  $h(\Si^{\om})$, for some ordinal $\alpha\geq 2$,  
hence there exists 
a  ${\bf \Si}^0_\alpha$   (respectively, ${\bf \Pi}^0_\alpha$) subset $T$ of 
 $(\Si \cup\{A, B, 0\})^\om$ such that 
$h(\mathcal{L})$=$T \cap h(\Si^{\om})$. But $h(\Si^{\om})$ is a closed 
i.e. ${\bf \Pi}^0_1$-subset 
of $(\Si \cup\{A, B, 0\})^\om$ and the class of ${\bf \Si}^0_\alpha$ 
(respectively, ${\bf \Pi}^0_\alpha$) subsets of 
$(\Si \cup\{A, B, 0\})^\om$  is closed under finite intersection thus 
$h(\mathcal{L})$ is a ${\bf \Si}^0_\alpha$   (respectively, ${\bf \Pi}^0_\alpha$) 
subset of $(\Si \cup\{A, B, 0\})^\om$.  
 
\hs   Now  $h(\mathcal{L})$ $\cup  (h(\Si^{\om}))^-$ is the union of a 
${\bf \Si}^0_\alpha$   (respectively, ${\bf \Pi}^0_\alpha$)         subset 
and of a ${\bf \Si}^0_1$-subset of $(\Si \cup\{A, B, 0\})^\om$ 
therefore it is a ${\bf \Si}^0_\alpha$     (respectively, ${\bf \Pi}^0_\alpha$)      
subset of $(\Si \cup\{A, B, 0\})^\om$ 
because the class of 
${\bf \Si}^0_\alpha$   (respectively, ${\bf \Pi}^0_\alpha$)  
 subsets of $(\Si \cup\{A, B, 0\})^\om$ 
 is closed under finite  union.  

  \hs Assume now firstly that $\mathcal{L}$ is  ${\bf \Si}^0_\alpha$-complete 
 (respectively, ${\bf \Pi}^0_\alpha$-complete). 
 In order to prove that $h(\mathcal{L})$$ \cup  (h(\Si^{\om}))^-$ 
 is ${\bf \Si}^0_\alpha$-{\bf complete}  
(respectively,   ${\bf \Pi}^0_\alpha$-{\bf complete})  
it suffices to remark that
\begin{center}
$\mathcal{L}$=$h^{-1}[ h(\mathcal{L}) $$\cup  (h(\Si^{\om}))^- ]$
\end{center}
\noi This implies that $h(\mathcal{L}) $$\cup  (h(\Si^{\om}))^- $ is 
${\bf \Si}^0_\alpha$-complete (respectively, ${\bf \Pi}^0_\alpha$-complete)    
because  $\mathcal{L}$ is assumed to be ${\bf \Si}^0_\alpha$-complete 
 (respectively, ${\bf \Pi}^0_\alpha$-complete).   

\hs On the other hand if we assume only that $\mathcal{L}$ is  a ${\bf \Si}^0_\alpha$-set of rank $\alpha$ 
 (respectively, ${\bf \Pi}^0_\alpha$-set of rank $\alpha$), then we can infer that $h(\mathcal{L})$$ \cup  (h(\Si^{\om}))^-$ 
is also a  ${\bf \Si}^0_\alpha$-set of rank $\alpha$ 
 (respectively, ${\bf \Pi}^0_\alpha$-set of rank $\alpha$). Indeed if $h(\mathcal{L})$$ \cup  (h(\Si^{\om}))^-$
was of Borel rank $\beta < \alpha$ then $\mathcal{L}$=$h^{-1}[ h(\mathcal{L}) $$\cup  (h(\Si^{\om}))^- ]$ would be 
also of rank smaller than $\alpha$ because the class ${\bf \Si}^0_\beta$ (respectively, ${\bf \Pi}^0_\beta$) is closed under 
inverse images by continuous functions.
 \ep  \end{proof}

\hs In order to apply Lemma \ref{lem3}, we want now to prove that 
if $L(\mathcal{A})$$\subseteq \Sio$ is accepted by a 
real time $2$-counter 
automaton $\mathcal{A}$ with a B\"uchi acceptance condition then 
$h( L(\mathcal{A}) )$$ \cup h(\Si^{\om})^-$ is accepted by a $1$-counter 
automaton with a B\"uchi acceptance condition.  
We firstly prove the following lemma.

\begin{Lem}\label{lem4}
Let $\Si$ be a finite alphabet and $h$ be the coding of $\om$-words over $\Si$ defined as 
above. Then $h(\Si^{\om})^-=(\Si\cup\{A, B, 0\})^\om - h(\Si^{\om})$ is accepted by a 
real time $1$-counter B\"uchi automaton. 
\end{Lem}

\begin{proof} 
 We can easily see  that $h(\Si^{\om})^-=(\Si\cup\{A, B, 0\})^\om - h(\Si^{\om})$ 
is the set of $\om$-words in 
$(\Si\cup\{A, B, 0\})^\om $ which belong to one of the following $\om$-languages. 

\begin{itemize} 
\ite $\mathcal{D}$$_1$ is the set of $\om$-words over the alphabet $\Si\cup\{A, B, 0\}$ 
which have not any initial segment in  $A.0^6.\Si.B$. It is easy to see that 
$\mathcal{D}$$_1$ is in fact a regular $\om$-language. 

\ite $\mathcal{D}$$_2$ is the complement of $(A.0^+.\Si.B.0^+)^\omega$ in 
$(\Si\cup\{A, B, 0\})^\om$. The $\om$-language $(A.0^+.\Si.B.0^+)^\omega$ is regular 
thus its complement $\mathcal{D}$$_2$ is also a regular $\om$-language.

\ite $\mathcal{D}$$_3$ is the set of $\om$-words over the alphabet $\Si\cup\{A, B, 0\}$ 
which contain a segment in  $B.0^n.A.0^m.\Si$ for some positive integers $n\neq m$. 
It is easy to see that this $\om$-language can be accepted by a 
real time $1$-counter B\"uchi automaton. 

\ite $\mathcal{D}$$_4$ is the set of $\om$-words over the alphabet $\Si\cup\{A, B, 0\}$ 
which contain a segment in  $A.0^n.\Si.B.0^m.A$ for some positive integers $n$ and $m$
with $m\neq 6n$. Again this $\om$-language can be accepted by a 
real time $1$-counter B\"uchi automaton. 
\end{itemize}

\noi 
The class {\bf r}-${\bf BCL}(1)_\om$ is closed under finite union because it is the class of 
$\om$-languages accepted by {\it non deterministic} real time $1$-counter B\"uchi automata. 
On the other hand it holds that 
$h(\Si^{\om})^-=(\Si\cup\{A, B, 0\})^\om - h(\Si^{\om}) = \cup_{1\leq i\leq 4} 
\mathcal{D}$$_i$ 
thus $h(\Si^{\om})^-$ is accepted by a real time 
$1$-counter B\"uchi automaton.
\ep  \end{proof}

\hs
We would like now to prove that 
if $L(\mathcal{A})$$\subseteq \Sio$ is accepted by a 
real time $2$-counter 
automaton $\mathcal{A}$ with a B\"uchi acceptance condition then 
$h( L(\mathcal{A}) )$ is in ${\bf BCL}(1)_\om$. 
We cannot show this, so we are firstly going to define another 
$\om$-language $\mathcal{L}$$(\mathcal{A})$ accepted by 
a $1$-counter B\"uchi automaton and we shall prove 
that 
$h( L(\mathcal{A}) ) $$\cup h(\Si^{\om})^- = \mathcal{L}$$(\mathcal{A})$$ \cup h(\Si^{\om})^-$. 
\nl  We shall need the following notion. Let $N \geq 1$ be an integer such that 
$N=2^x.3^y.N_1$ where $x, y$ are positive integers and $N_1 \geq 1$ is an integer which is 
neither divisible by $2$ nor by $3$. Then we set $P_2(N)=x$ and $P_3(N)=y$. So $2^{P_2(N)}$ is 
the greatest power of $2$ which divides $N$ and $2^{P_3(N)}$ is 
the greatest power of $3$ which divides $N$.  
\nl  Let then a $2$-counter B\"uchi automaton $\mathcal{A}$$=(K,\Si, \Delta, q_0, F)$ 
accepting the $\om$-language  $L(\mathcal{A})$$\subseteq \Sio$.
 The $\om$-language 
$\mathcal{L}$$(\mathcal{A})$ is the set of $\om$-words over the alphabet 
$\Si \cup\{A, B, 0\}$ in the form 
$$A.u_1.v_1.x_1.B.w_1.z_1.A.u_2.v_2.x_2.B.w_2.z_2.A \ldots 
A.u_n.v_n.x_n.B.w_n.z_n.A \ldots $$

\noi where,  for all integers $i\geq 1$, $v_i, w_i \in 0^+$, 
 $u_i, z_i \in 0^\star$, $x_i\in \Si$, $|u_1|=5$, 
$|u_{i+1}|=|z_i|$ and there is a sequence $(q_i)_{i\geq 0}$  of states of $K$ and  
integers $j_i, j'_i \in \{-1; 0; 1\}$, for $i\geq 1$, such that  for all integers 
$i\geq 1$:  

$$  x_i : ( q_{i-1}, P_2(|v_i|), P_3(|v_i|) ) \mapsto_{\mathcal{A}} 
(q_i, P_2(|v_i|) + j_i , P_3(|v_i|) + j'_i )$$

\noi and 
$$|w_i| = |v_i|.2^{j_i}.3^{j'_i}$$
\noi Moreover some state $q_f \in F$ occurs infinitely often in the sequence $(q_i)_{i\geq 0}$. 
\nl  Notice that the state $q_0$ of the sequence $(q_i)_{i\geq 0}$  is also the initial state 
of $\mathcal{A}$.

\begin{Lem}\label{lem4-}
 Let $\mathcal{A}$ be a real time 
$2$-counter B\"uchi automaton accepting 
$\om$-words over the alphabet $\Si$ and 
$\mathcal{L}$$(\mathcal{A})$$\subseteq (\Si \cup\{A, B, 0\})^\om$  be defined as above. 
Then $\mathcal{L}(\mathcal{A})$ is accepted by a $1$-counter B\"uchi automaton $\mathcal{B}$.
\end{Lem}

\begin{proof}
 We shall explain informally the behaviour of 
a $1$-counter B\"uchi automaton $\mathcal{B}$ accepting the $\om$-language 
$\mathcal{L}(\mathcal{A})$. 
\nl We firstly 
consider the reading of an $\om$-word $x \in (A.0^\star.\Si.B.0^\star)^\om$ in the form
$$x = A.0^{n_1}x_1.B.0^{m_1}.A.0^{n_2}x_2.B.0^{m_2}.A \ldots A.0^{n_p}x_p.B.0^{m_p}.A \ldots $$
\noi where, for all integers $i\geq 1$, $n_i, m_i$,  are positive integers and  
$x_i \in \Si$. 
\nl Using the finite control the automaton $\mathcal{B}$  first checks that the six first 
letters of $x$ form the initial segment $A.0^5$. Then,  when reading the following $(n_1-5)$ 
letters $0$, the automaton $\mathcal{B}$, using the finite control, 
checks that $(n_1-5)>0$ and determines whether $P_2(n_1-5)=0$ and whether $P_3(n_1-5)=0$. 
Moreover the counter content is increased by one for each letter $0$ read. 
The automaton $\mathcal{B}$  reads now the letter $x_1$ and it guesses a transition 
of $\mathcal{A}$ leading to  
$$x_1 : ( q_{0}, P_2(n_1-5), P_3(n_1-5) ) \mapsto_{\mathcal{A}} 
(q_1, P_2(n_1-5) + j_1 , P_3(n_1-5) + j'_1 )$$
\noi We set $v_1=0^{n_1-5}$ and $w_1=0^{(n_1-5).2^{j_1}.3^{j'_1}}$. 
The counter value is now equal to $(n_1-5)$ and, when reading letters $0$ following $x_1$, 
the automaton $\mathcal{B}$ checks that $m_1 \geq (n_1-5).2^{j_1}.3^{j'_1}$ in such a way 
that the counter value becomes $0$ after having read the  $(n_1-5).2^{j_1}.3^{j'_1}$ letters 
$0$ following the first letter $B$. For instance if $j_1=j'_1=1$ then $|w_1|=|v_1|.6$ so this 
can be done by decreasing the counter content by one each time six letters $0$ are read. 
The other cases are treated in a similar way. Details are here left to the reader.
\nl Notice also that  the automaton $\mathcal{B}$ has kept in its finite control the 
value of the state $q_1$. 
\nl We set now $0^{m_1}=w_1.z_1$. We have seen that after having read $w_1$ the counter value 
is equal to zero. Now when reading $z_1$ the counter content is increased by one for each letter 
read so that it becomes $|z_1|$ after having read $z_1$. The automaton $\mathcal{B}$  
reads now a letter $A$ and next decreased its counter by one for each letter $0$ read until the 
counter content is equal to zero. We set $0^{n_2}=u_2.v_2 $ with $u_2=z_1$. The automaton 
$\mathcal{B}$ reads now the segment $v_2$. 
Using the finite control, it 
checks that $|v_2|>0$ and determines whether $P_2(|v_2|)=0$ and whether $P_3(|v_2|)=0$. 
Moreover the counter content is increased by one for each letter $0$ read. 
The automaton $\mathcal{B}$  reads now the letter $x_2$ and it guesses a transition 
of $\mathcal{A}$ leading to  
$$x_2 : ( q_{1}, P_2(|v_2|), P_3(|v_2|) ) \mapsto_{\mathcal{A}} 
(q_2, P_2(|v_2|) + j_2 , P_3(|v_2|) + j'_2 )$$
\noi We set  $w_2=0^{|v_2|.2^{j_2}.3^{j'_2}}$. 
The counter value is now equal to $|v_2|$. The automaton 
 $\mathcal{B}$  reads now  the second letter $B$ and, when reading the $m_2$ 
letters $0$ following this letter $B$, 
the automaton $\mathcal{B}$ checks that $m_2 \geq |v_2|.2^{j_2}.3^{j'_2}$ in such a way 
that the counter value becomes $0$ after having read the  $|v_2|.2^{j_2}.3^{j'_2}$ letters 
$0$ following the second  letter $B$. 
\nl For instance if $j_2=0$ and $j'_2=-1$ 
then $|w_2|=|v_2|.3^{-1}$ so this 
can be done by decreasing the counter content by three each time one letter $0$ is read. 
\nl And if $j_2=-1$ and $j'_2=-1$  then $|w_2|=|v_2|.2^{-1}.3^{-1}=|v_2|.6^{-1}$ so this 
can be done by decreasing the counter content by six each time one letter $0$ is read. 
The other cases are treated in a similar way. Details are here left to the reader.
\nl Notice that these different cases 
 can be achieved with the use of  $\lambda$-transitions but in such a way 
that there will be at most 5 consecutive $\lambda$-transitions during a run of $\mathcal{B}$
on $x$. This will be an important useful fact in the sequel. 
\nl Notice also that  the automaton $\mathcal{B}$ has kept in its finite control the 
value of the state $q_2$. 
\nl The reading of $x$ by $\mathcal{B}$ continues in the same way. A B\"uchi acceptance 
condition  can be used to ensure that some state $q_f\in K$ 
occurs infinitely often in the sequence $(q_i)_{i\geq 0}$. 
\nl To complete the proof we can remark that $\mathcal{R}$=$(A.0^\star.\Si.B.0^\star)^\om$ is a 
regular $\om$-language so we have considered only the reading by $\mathcal{B}$ of $\om$-words 
$x\in \mathcal{R}$. Indeed if the $\om$-language $L(\mathcal{B})$ was not included into 
$\mathcal{R}$ we could replace it by $L(\mathcal{B})$$\cap \mathcal{R}$ because the class 
${\bf BCL}(1)_\om$ is closed under intersection with regular $\om$-languages (by a classical 
 construction of  product of automata, the  $\om$-language $\mathcal{R}$ being 
accepted by a deterministic Muller automaton).  
\ep  \end{proof}

\begin{Lem}\label{lem5}  Let $\mathcal{A}$ be a real time 
$2$-counter B\"uchi automaton accepting 
$\om$-words over the alphabet $\Si$ and 
$\mathcal{L}(\mathcal{A})$$\subseteq (\Si \cup\{A, B, 0\})^\om$  be defined as above. Then 
$ L(\mathcal{A})$ = $h^{-1} (\mathcal{L}(\mathcal{A}))$, i.e. 
$\fa x \in \Si^{\om} ~~~~ h(x) \in \mathcal{L}(\mathcal{A}) $$\longleftrightarrow 
x\in  L(\mathcal{A})$. 
\end{Lem}

\begin{proof}
 Let $\mathcal{A}$ be a real time 
$2$-counter B\"uchi automaton accepting 
$\om$-words over the alphabet $\Si$ and 
$\mathcal{L}(\mathcal{A})$$\subseteq (\Si \cup\{A, B, 0\})^\om$  be defined as above. 
Let $x\in \Si^{\om}$ be an $\om$-word such that $h(x) \in \mathcal{L}(\mathcal{A})$. 
So  $h(x)$ may  be written 
$$h(x)=A.0^6.x(1).B.0^{6^2}.A.0^{6^2}.x(2).B.0^{6^3}.A.0^{6^3}.x(3).B \ldots  
B.0^{6^n}.A.0^{6^n}.x(n).B \ldots  $$
\noi and also 
$$h(x)=A.u_1.v_1.x_1.B.w_1.z_1.A.u_2.v_2.x_2.B.w_2.z_2.A \ldots 
A.u_n.v_n.x_n.B.w_n.z_n.A \ldots $$

\noi where,  for all integers $i\geq 1$, $v_i, w_i \in 0^+$, 
 $u_i, z_i \in 0^\star$, $x_i=x(i) \in \Si$, $|u_1|=5$, 
$|u_{i+1}|=|z_i|$ and there is a sequence $(q_i)_{i\geq 0}$  of states of $K$ and  
integers $j_i, j'_i \in \{-1; 0; 1\}$, for $i\geq 1$, such that  for all integers 
$i\geq 1$:  

$$  x_i : ( q_{i-1}, P_2(|v_i|), P_3(|v_i|) ) \mapsto_{\mathcal{A}} 
(q_i, P_2(|v_i|) + j_i , P_3(|v_i|) + j'_i )$$

\noi and 
$$|w_i| = |v_i|.2^{j_i}.3^{j'_i}$$
\noi some state $q_f \in F$ occurring infinitely often in the sequence $(q_i)_{i\geq 0}$. 
\nl  In particular, $u_1=0^5$ and $u_1.v_1=0^6$ thus $|v_1|=1=2^0.3^0$. 
We are going to prove by induction on the integer $i\geq 1$ that, 
for all integers $i\geq 1$, 
$|w_i|=|v_{i+1}|=2^{P_2(|w_i|)}.3^{P_3(|w_i|)}$. Moreover, setting $c_1^{i}=P_2(|v_i|)$ and 
$c_2^{i}=P_3(|v_i|)$, we are going to prove that for each integer $i\geq 1$ it holds that 
$$  x_i : ( q_{i-1}, c_1^{i}, c_2^{i} ) \mapsto_{\mathcal{A}} 
(q_i,  c_1^{i+1},  c_2^{i+1} )$$

\noi We have already seen that $|v_1|=1=2^0.3^0$. By hypothesis there is a state 
$q_1 \in K$ and integers $j_1, j'_1 \in \{-1; 0; 1\}$ such that 
$ x_1 : ( q_{0}, P_2(|v_1|), P_3(|v_1|) ) \mapsto_{\mathcal{A}} 
(q_1, P_2(|v_1|) + j_1 , P_3(|v_1|) + j'_1 )$, i.e. 
$ x_1 : ( q_{0}, 0, 0 ) \mapsto_{\mathcal{A}} 
(q_1,  j_1 ,  j'_1 )$. Then $|w_1|=|v_1|.2^{j_1}.3^{j'_1}=2^{j_1}.3^{j'_1}$. 
\nl We have now $|w_1.z_1|=|u_2.v_2|=0^{6^2}$ and $|u_2|=|z_1|$ thus 
$|v_2|=|w_1|=2^{j_1}.3^{j'_1}$. Setting $c_1^1=0$,  $c_2^1=0$,  
$c_1^2=j_1=P_2(|v_2|)$ and  $c_2^2=j'_1=P_3(|v_2|)$, it holds that 
$x_1 : ( q_{0}, c_1^1, c_2^1 ) \mapsto_{\mathcal{A}} 
(q_1,   c_1^2,  c_2^2 )$.   
\nl 
 Assume now that, for all integers $i$, $1\leq i \leq n-1$, it holds that 
$|w_i|=|v_{i+1}|=2^{P_2(|w_i|)}.3^{P_3(|w_i|)}$ and 
$x_i : ( q_{i-1}, c_1^{i}, c_2^{i} ) \mapsto_{\mathcal{A}} 
(q_i,  c_1^{i+1},  c_2^{i+1} )$ where 
 $c_1^{i}=P_2(|v_i|)$  and 
$c_2^{i}=P_3(|v_i|)$. 
\nl  
We know that there is a state $q_n\in K$ and integers $j_n, j'_n \in \{-1; 0; 1\}$ such that 
$ x_n : ( q_{n-1}, P_2(|v_n|), P_3(|v_n|) ) \mapsto_{\mathcal{A}} 
(q_n, P_2(|v_n|) + j_n , P_3(|v_n|) + j'_n )$, i.e. 
$ x_n : ( q_{n-1}, c_1^n, c_2^n ) \mapsto_{\mathcal{A}} 
(q_n,  c_1^n + j_n , c_2^n + j'_n )$. Then 
$|w_n|=|v_n|.2^{j_n}.3^{j'_n}=2^{c_1^n + j_n}.3^{c_2^n + j'_n}$. 
\nl On the other hand $|w_n.z_n|=|u_{n+1}.v_{n+1}|=0^{6^{n+1}}$ and $|u_{n+1}|=|z_n|$ thus 
$|v_{n+1}|=|w_n|=2^{c_1^n + j_n}.3^{c_2^n + j'_n}=2^{c_1^{n + 1}}.3^{c_2^{n + 1}}$ by setting 
 $c_1^{n + 1}=P_2(|v_{n+1}|)$ and $c_2^{n + 1}=P_3(|v_{n+1}|)$. So we have 
$x_{n} : ( q_{n-1}, c_1^n, c_2^n ) \mapsto_{\mathcal{A}} 
(q_n,   c_1^{n + 1},  c_2^{n + 1})$. 

\hs Finally we have proved by induction the announced claim. If for all integers 
$i\geq 1$, we set  $c_1^{i}=P_2(|v_i|)$ and 
$c_2^{i}=P_3(|v_i|)$,  it holds that 
$$  x_i : ( q_{i-1}, c_1^{i}, c_2^{i} ) \mapsto_{\mathcal{A}} 
(q_i,  c_1^{i+1},  c_2^{i+1} )$$

\noi But there is some state $q_f\in K$ which occurs infinitely often 
in the sequence $(q_i)_{i\geq 1}$. 
This  implies that $( q_{i-1}, c_1^{i}, c_2^{i} )_{i\geq 1}$ is a successful 
run of $\mathcal{A}$ on $x$ thus $x\in L(\mathcal{A})$. 
\nl  Conversely it is easy to see that if $x\in L(\mathcal{A})$ then 
$h(x)\in \mathcal{L}(\mathcal{A})$. This ends the proof of Lemma \ref{lem5}. 
\ep  \end{proof}

\begin{Rem}
The simulation, during the reading of $h(x)$ by the  
$1$-counter B\"uchi automaton $\mathcal{B}$,  of the behaviour of the real time 
$2$-counter B\"uchi automaton  $\mathcal{A}$ reading  $x$, 
 can be achieved, using a coding of the content $(c_1, c_2)$ of two counters 
by a single integer $2^{c_1}.3^{c_2}$ and the {\bf special shape} of $\om$-words 
in $h(\Sio)$ which allows the propagation of the counter value of $\mathcal{B}$. 
 This will be sufficient here, because of the  previous lemmas, and in particular of the fact 
that $h(\Sio)^-$ is in the class {\bf r}-${\bf BCL}(1)_\om$,  
 and we can now end the 
proof of Proposition \ref{prop}. 
\end{Rem}

{\it End of Proof of Proposition \ref{prop}.} 
Let  $\alpha \geq 2$ be an ordinal. Assume that  
there is an 
$\om$-language $L(\mathcal{A})$$\subseteq \Sio$  which is  
${\bf \Si}^0_\alpha$-complete (respectively,   ${\bf \Pi}^0_\alpha$-complete,  
${\bf \Si}^0_\alpha$ of rank $\alpha$, ${\bf \Pi}^0_\alpha$ of rank $\alpha$) and is 
accepted by a 
 real time $2$-counter B\"uchi automaton  $\mathcal{A}$. 
By Lemma \ref{lem3}, $h(\mathcal{L})$$ \cup h(\Si^{\om})^-$
 is a 
subset of $(\Si \cup\{A, B, 0\})^\om$ being ${\bf \Si}^0_\alpha$-complete 
(respectively, ${\bf \Pi}^0_\alpha$-complete, ${\bf \Si}^0_\alpha$ of rank $\alpha$, ${\bf \Pi}^0_\alpha$ of rank $\alpha$) . 
On the other hand Lemma \ref{lem5} states that 
$L(\mathcal{A})$ = $h^{-1} (\mathcal{L}(\mathcal{A}))$ and this implies that 
$h( L(\mathcal{A}) ) $$\cup h(\Si^{\om})^- = \mathcal{L}(\mathcal{A}) $$\cup h(\Si^{\om})^-$.
But we know by Lemmas \ref{lem4} and \ref{lem4-} that the $\om$-languages 
$h(\Si^{\om})^-$ and $\mathcal{L}(\mathcal{A})$ 
are in the class 
${\bf BCL}(1)_\om$ thus their union 
is also accepted by a $1$-counter 
B\"uchi automaton. 
Therefore  $h( L(\mathcal{A}) ) $$\cup h(\Si^{\om})^-$ is an $\om$-language 
in the class  ${\bf BCL}(1)_\om$ which is ${\bf \Si}^0_\alpha$-complete 
(respectively,   ${\bf \Pi}^0_\alpha$-complete, ${\bf \Si}^0_\alpha$ of rank $\alpha$, ${\bf \Pi}^0_\alpha$ of rank $\alpha$). 

\hs We want now  to find an $\om$-language 
in the class  {\bf r}-${\bf BCL}(1)_\om$ which is ${\bf \Si}^0_\alpha$-complete 
(respectively,   ${\bf \Pi}^0_\alpha$-complete, ${\bf \Si}^0_\alpha$ of rank $\alpha$, ${\bf \Pi}^0_\alpha$ of rank $\alpha$). 
\nl On one side we have proved that $h(\Si^{\om})^-$ is accepted by a {\it real time} 
$1$-counter  B\"uchi automaton. 
 On the other side we have proved that $\mathcal{L}(\mathcal{A})$  is accepted by a 
(non real time) $1$-counter  B\"uchi automaton $\mathcal{B}$. 
However we have seen, in the proof of Lemma 
\ref{lem4-}, that at most 5 consecutive $\lambda$-transitions can occur during the reading 
of an $\om$-word $x$ by  $\mathcal{B}$. 
\nl Consider now the mapping 
$\phi : (\Si\cup\{A, B, 0\})^\om \ra (\Si\cup\{A, B, F,  0\})^\om $ which is  defined, 
for all $x\in (\Si\cup\{A, B, 0\})^\om$ by: 
$$\phi(x) = F^5.x(1).F^5.x(2).F^5.x(3) \ldots F^5.x(n). F^5.x(n+1).F^5 \ldots$$
\noi The function $\phi$ is continuous and separates two successive letters of $x$ by 
five letters $F$. We can prove, as in the proof of Lemma \ref{thetaborel}, that if 
$L \subseteq (\Si\cup\{A, B, 0\})^\om$ is ${\bf \Si}^0_\alpha$-complete 
(respectively,   ${\bf \Pi}^0_\alpha$-complete, ${\bf \Si}^0_\alpha$ of rank $\alpha$, ${\bf \Pi}^0_\alpha$ of rank $\alpha$), 
for some ordinal $\alpha \geq 2$, then 
$\phi(L)$ is a  subset of $(\Si\cup\{A, B, F,  0\})^\om $ which is ${\bf \Si}^0_\alpha$-complete 
(respectively,   ${\bf \Pi}^0_\alpha$-complete,  ${\bf \Si}^0_\alpha$ of rank $\alpha$, ${\bf \Pi}^0_\alpha$ of rank $\alpha$). 

\hs  Thus the $\om$-language $\phi(  \mathcal{L}(\mathcal{A})$$ \cup h(\Si^{\om})^- )$ is 
${\bf \Si}^0_\alpha$-complete 
(respectively,   ${\bf \Pi}^0_\alpha$-complete, ${\bf \Si}^0_\alpha$ of rank $\alpha$, ${\bf \Pi}^0_\alpha$ of rank $\alpha$). 
\nl Moreover it is easy to see that $\phi(  \mathcal{L}(\mathcal{A}) )$ 
is accepted by a {\it real time} $1$-counter 
B\"uchi  automaton $\mathcal{B}'$. The automaton $\mathcal{B}'$ checks with its finite control 
that an input $\om$-word is in the form $\phi(x)$ for some $x\in (\Si\cup\{A, B, 0\})^\om$. 
And $\mathcal{B}'$ simulates the reading of $x$ by $\mathcal{B}$, the $\lambda$-transitions 
of $\mathcal{B}$ occuring during the reading, in {\it real time},  of letters $F$ of the 
$\om$-word $\phi(x)$. 
\nl Finally $\phi(  \mathcal{L}(\mathcal{A}) $$\cup h(\Si^{\om})^- ) = 
\phi(  \mathcal{L}(\mathcal{A}) )  $$\cup \phi(  h(\Si^{\om})^- ) )$ is the union of 
two $\om$-languages in {\bf r}-${\bf BCL}(1)_\om$ thus it is in {\bf r}-${\bf BCL}(1)_\om$ 
and it is ${\bf \Si}^0_\alpha$-complete 
(respectively,   ${\bf \Pi}^0_\alpha$-complete, ${\bf \Si}^0_\alpha$ of rank $\alpha$, ${\bf \Pi}^0_\alpha$ of rank $\alpha$).
\nl This ends the proof of Proposition \ref{prop} for the integer $k=2$. 

\hs We explain now the modifications to do in order to prove Proposition \ref{prop} 
for the integer $k=8$. We assume that $\alpha \geq 2$ is an ordinal and  that  
there is an  
$\om$-language $L(\mathcal{A})$$\subseteq \Sio$  which is 
${\bf \Si}^0_\alpha$-complete (respectively,   ${\bf \Pi}^0_\alpha$-complete,  ${\bf \Si}^0_\alpha$ of rank $\alpha$, ${\bf \Pi}^0_\alpha$ of rank $\alpha$) 
and is accepted by a 
 real time $8$-counter B\"uchi automaton  $\mathcal{A}$. 
\nl We first modify the coding  of $\om$-words which was given by the mapping $h$. 
We replace the number $6=2.3$ by the product of the eight first prime numbers: 
$$K = 2.3.5.7.11.13.17.19 = 9699690$$
\noi Then an $\om$-word $x\in \Sio$ will be coded by the $\om$-word 
$$h_K(x)=A.0^K.x(1).B.0^{K^2}.A.0^{K^2}.x(2).B.0^{K^3}.A.0^{K^3}.x(3).B \ldots  
B.0^{K^n}.A.0^{K^n}.x(n).B \ldots  $$
\noi The mapping $h_K : \Sio \ra  (\Si \cup\{A, B, 0\})^\om$  is continuous and 
we can prove, as in Lemma \ref{lem3}, that 
$h_K(L(\mathcal{A}))$$ \cup h_K(\Si^{\om})^-$
is a  subset of $(\Si \cup\{A, B, 0\})^\om$ which is 
${\bf \Si}^0_\alpha$-complete (respectively, ${\bf \Pi}^0_\alpha$-complete, ${\bf \Si}^0_\alpha$ of rank $\alpha$, ${\bf \Pi}^0_\alpha$ of rank $\alpha$).
 As in Lemma \ref{lem4}, we can prove that  $h_K(\Si^{\om})^-$ is in the class 
{\bf r}-${\bf BCL}(1)_\om$. 
\nl Next, for each prime number $p\in \{2; 3; 5; 7; 11; 13; 17; 19\}$, and 
each positive integer $N\geq 1$, we denote $P_p(N)$ the positive integer such that 
$p^{P_p(N)}$ is the greatest power of $p$ which divides $N$. 
\nl We  define the  $\om$-language 
$\mathcal{L}(\mathcal{A})$ as  the set of $\om$-words over the alphabet 
$\Si \cup\{A, B, 0\}$ in the form 
$$A.u_1.v_1.x_1.B.w_1.z_1.A.u_2.v_2.x_2.B.w_2.z_2.A \ldots 
A.u_n.v_n.x_n.B.w_n.z_n.A \ldots $$

\noi where,  for all integers $i\geq 1$, $v_i, w_i \in 0^+$, 
 $u_i, z_i \in 0^\star$, $|u_1|=K-1$, 
$|u_{i+1}|=|z_i|$ and there is a sequence $(q_i)_{i\geq 0}$  of states of $K$ and  
integers $j_i^1, j_i^2, \ldots, j_i^8,  \in \{-1; 0; 1\}$, 
for $i\geq 1$, such that  for all integers 
$i\geq 1$:  

$$  x_i : ( q_{i-1}, P_2(|v_i|), P_3(|v_i|), \ldots, P_{19}(|v_i|) ) \mapsto_{\mathcal{A}} 
(q_i, P_2(|v_i|) + j_i^1 , P_3(|v_i|) + j_i^2, \ldots,  P_{19}(|v_i|) + j_i^8 )$$

\noi and 
$$|w_i| = |v_i|.2^{j_i^1}.3^{j_i^2}. \ldots .19^{j_i^8}$$
\noi and some state $q_f \in F$ occurs infinitely often in the sequence $(q_i)_{i\geq 0}$.

\hs Applying the same  ideas as in the proofs of Lemmas \ref{lem4-} and 
 \ref{lem5} we can prove that $\mathcal{L}(\mathcal{A})$ is accepted by a 
$1$-counter B\"uchi automaton and that $L(\mathcal{A}) $$= h_K^{-1} (\mathcal{L}(\mathcal{A}))$. 
\nl The essential change is that now  the content $(c_1, c_2, \ldots, c_8)$ 
of eight counters is coded by the product $2^{c_1}.3^{c_2}. \ldots .(17)^{c_7}.(19)^{c_8}$. 
\nl Notice that again  $\mathcal{L}(\mathcal{A})$  is accepted by a 
(non real time) $1$-counter  B\"uchi automaton $\mathcal{B}$.  
However there are now 
at most $(K-1)$ consecutive $\lambda$-transitions which can  occur during the reading 
of an $\om$-word $x$ by  $\mathcal{B}$.
\nl So we  define now the mapping 
$\phi_K : (\Si\cup\{A, B, 0\})^\om \ra (\Si\cup\{A, B, F,  0\})^\om $ by:  
for all $x\in (\Si\cup\{A, B, 0\})^\om$, 
$$\phi_K(x) = F^{K-1}.x(1).F^{K-1}.x(2).F^{K-1}.x(3) 
\ldots F^{K-1}.x(n). F^{K-1}.x(n+1).F^{K-1} \ldots$$
\noi The function $\phi_K$ is continuous as the function $\phi$ was.  
  The end of the proof is unchanged so we infer that 
$\phi_K ( h_K( L(\mathcal{A}) )$$ \cup h_K(\Si^{\om})^- )$ is an $\om$-language 
in the class  {\bf r}-${\bf BCL}(1)_\om$ which is 
${\bf \Si}^0_\alpha$-complete 
(respectively,   ${\bf \Pi}^0_\alpha$-complete, ${\bf \Si}^0_\alpha$ of rank $\alpha$, ${\bf \Pi}^0_\alpha$ of rank $\alpha$).  
\ep

\hs From the results of  Section \ref{section4} and Proposition \ref{prop}, we can now 
 state the following  result.

\begin{The}\label{thebor}    

\noi Let $\mathcal{C}$ be a class of $\om$-languages such that:  
\begin{center}
$\mbox{ {\bf r}-}{\bf BCL}(1)_\om \subseteq  \mathcal{C}$$  \subseteq  \Sigma^1_1.$
\end{center}
\begin{enumerate} 
\ite[(a)]  The Borel hierarchy of the class $\mathcal{C}$ is equal to the Borel hierarchy of the class $\Sigma^1_1$. 
\ite[(b)]   $\gamma_2^1=Sup ~~ \{ \alpha \mid \exists L \in \mathcal{C} \mbox{ such that }$$L $$\mbox{ is a Borel set of rank } \alpha \}.$  
\ite[(c)]  For every non null  ordinal $\alpha < \om_1^{\mathrm{CK}}$, 
there exists some  
${\bf \Si}^0_\alpha$-complete and some ${\bf \Pi}^0_\alpha$-complete
$\om$-languages in the class $\mathcal{C}$.  
\end{enumerate}
\end{The}

\noi Notice that above $(b)$ and $(c)$ just follow from $(a)$ and from the known results about the Borel hierarchy of the class $\Sigma^1_1$.

\section{Wadge hierarchy of  $\om$-languages in {\bf r}-${\bf BCL}(1)_\om$}\label{section-wadge}

\noi We now introduce the Wadge hierarchy, which is a great refinement of the Borel hierarchy defined 
via reductions by continuous functions, \cite{Duparc01,Wadge83}. 

\begin{Deff}[Wadge \cite{Wadge83}] Let $X$, $Y$ be two finite alphabets. 
For $L\subseteq X^\om$ and $L'\subseteq Y^\om$, $L$ is said to be Wadge reducible to $L'$
($L\leq _W L')$ iff there exists a continuous function $f: X^\om \ra Y^\om$, such that
$L=f^{-1}(L')$.
\nl $L$ and $L'$ are Wadge equivalent iff $L\leq _W L'$ and $L'\leq _W L$. 
This will be denoted by $L\equiv_W L'$. And we shall say that 
$L<_W L'$ iff $L\leq _W L'$ but not $L'\leq _W L$.
\nl  A set $L\subseteq X^\om$ is said to be self dual iff  $L\equiv_W L^-$, and otherwise 
it is said to be non self dual.
\end{Deff}

\noi
 The relation $\leq _W $  is reflexive and transitive,
 and $\equiv_W $ is an equivalence relation.
\nl The {\it equivalence classes} of $\equiv_W $ are called {\it Wadge degrees}. 
\nl The Wadge hierarchy $WH$ is the class of Borel subsets of a set  $X^\om$, where  $X$ is a finite set,
 equipped with $\leq _W $ and with $\equiv_W $.
\nl  For $L\subseteq X^\om$ and $L'\subseteq Y^\om$, if   
$L\leq _W L'$ and $L=f^{-1}(L')$  where $f$ is a continuous 
function from $ X^\om$  into $Y^\om$, then $f$ is called a continuous reduction of $L$ to 
$L'$. Intuitively it means that $L$ is less complicated than $L'$ because 
to check whether $x\in L$ it suffices to check whether $f(x)\in L'$ where $f$ 
is a continuous function. Hence the Wadge degree of an \ol~
is a measure 
of its topological complexity. 
\nl
Notice  that in the above definition, we consider that a subset $L\subseteq  X^\om$ is given
together with the alphabet $X$.

\noi We can now define the {\it Wadge class} of a set $L$:

\begin{Deff}
Let $L$ be a subset of $X^\om$. The Wadge class of $L$ is :
$$[L]= \{ L' \mid  L'\subseteq Y^\om \mbox{ for a finite alphabet }Y   \mbox{  and  } L'\leq _W L \}.$$ 
\end{Deff}

\noi Recall that each {\it Borel class} ${\bf \Si^0_\alpha}$ and ${\bf \Pi^0_\alpha}$ 
is a {\it Wadge class}. 
A set $L\subseteq X^\om$ is a ${\bf \Si^0_\alpha}$
 (respectively ${\bf \Pi^0_\alpha}$)-{\it complete set} iff for any set 
$L'\subseteq Y^\om$, $L'$ is in 
${\bf \Si^0_\alpha}$ (respectively ${\bf \Pi^0_\alpha}$) iff $L'\leq _W L $ .
 
\hs  There is a close relationship between Wadge reducibility
 and games which we now introduce.  

\begin{Deff} Let 
$L\subseteq X^\om$ and $L'\subseteq Y^\om$. 
The Wadge game  $W(L, L')$ is a game with perfect information between two players,
player 1 who is in charge of $L$ and player 2 who is in charge of $L'$.
\nl Player 1 first writes a letter $a_1\in X$, then player 2 writes a letter
$b_1\in Y$, then player 1 writes a letter $a_2\in  X$, and so on. 
\nl The two players alternatively write letters $a_n$ of $X$ for player 1 and $b_n$ of $Y$
for player 2.
\nl After $\om$ steps, the player 1 has written an $\om$-word $a\in X^\om$ and the player 2
has written an $\om$-word $b\in Y^\om$.
 The player 2 is allowed to skip, even infinitely often, provided he really writes an
$\om$-word in  $\om$ steps.
\nl The player 2 wins the play iff [$a\in L \lra b\in L'$], i.e. iff : 
\begin{center}
  [($a\in L ~{\rm and} ~ b\in L'$)~ {\rm or} ~ 
($a\notin L ~{\rm and}~ b\notin L'~{\rm and} ~ b~{\rm is~infinite}  $)].
\end{center}
\end{Deff}

\noi
Recall that a strategy for player 1 is a function 
$\sigma :(Y\cup \{s\})^\star\ra X$.
And a strategy for player 2 is a function $f:X^+\ra Y\cup\{ s\}$.
\nl $\sigma$ is a winning stategy  for player 1 iff he always wins a play when
 he uses the strategy $\sigma$, i.e. when the  $n^{th}$  letter he writes is given
by $a_n=\sigma (b_1\ldots b_{n-1})$, where $b_i$ is the letter written by player 2 
at step $i$ and $b_i=s$ if player 2 skips at step $i$.
\nl A winning strategy for player 2 is defined in a similar manner.

\hs   Martin's Theorem states that every Gale-Stewart Game $G(X)$ (see \cite{Kechris94}),  with $X$ a borel set, 
is determined and this implies the following :

\begin{The} [Wadge] Let $L\subseteq X^\om$ and $L'\subseteq Y^\om$ be two Borel sets, where
$X$ and $Y$ are finite  alphabets. Then the Wadge game $W(L, L')$ is determined :
one of the two players has a winning strategy. And $L\leq_W L'$ iff the player 2 has a 
winning strategy  in the game $W(L, L')$.
\end{The}

\begin{The} [Wadge]\label{wh}
Up to the complement and $\equiv _W$, the class of Borel subsets of $X^\om$,
 for  a finite alphabet $X$, is a well ordered hierarchy.
 There is an ordinal $|WH|$, called the length of the hierarchy, and a map
$d_W^0$ from $WH$ onto $|WH|-\{0\}$, such that for all $L, L' \subseteq X^\om$:
\nl $d_W^0 L < d_W^0 L' \lra L<_W L' $  and 
\nl $d_W^0 L = d_W^0 L' \lra [ L\equiv_W L' $ or $L\equiv_W L'^-]$.
\end{The}

\noi 
 The Wadge hierarchy of Borel sets of {\bf finite rank }
has  length $^1\varepsilon_0$ where $^1\varepsilon_0$
 is the limit of the ordinals $\alpha_n$ defined by $\alpha_1=\om_1$ and 
$\alpha_{n+1}=\om_1^{\alpha_n}$ for $n$ a non negative integer, $\om_1$
 being the first non countable ordinal. Then $^1\varepsilon_0$ is the first fixed 
point of the ordinal exponentiation of base $\om_1$. The length of the Wadge hierarchy 
of Borel sets in ${\bf \Delta^0_\om}= {\bf \Si^0_\om}\cap {\bf \Pi^0_\om}$ 
  is the $\om_1^{th}$ fixed point 
of the ordinal exponentiation of base $\om_1$, which is a much larger ordinal. The length 
of the whole Wadge hierarchy of Borel sets is a huge ordinal, with regard 
to the $\om_1^{th}$ fixed point 
of the ordinal exponentiation of base $\om_1$. It is described in \cite{Wadge83,Duparc01} 
by the use of the Veblen functions. 

\hs It is natural to ask for the Wadge hierarchy of classes of $\om$-languages accepted by 
finite machines, like {\bf X}-automata. 
The Wadge hierarchy of regular $\om$-languages, now called the Wagner hierarchy, has been 
effectively determined by Wagner; it has length $\om^\om$ \cite{Wagner79,Selivanov95,Selivanov98}. 
Wilke and Yoo  
proved in \cite{WilkeYoo95} that  one can compute in polynomial time the Wadge degree of an \orl. 
The Wadge hierarchy of \ol s 
accepted by Muller {\it deterministic}  one blind (i. e. without zero-test)  counter 
 automata is an effective extension of the Wagner hierarchy studied in \cite{Fin01csl}. 
Its extension to   {\it deterministic}                                   context free $\om$-languages has been determined by Duparc, its
 length is $\om^{(\om^2)}$ \cite{DFR,Duparc03} but we do not know yet 
whether it is  effective. 
Selivanov has recently determined the Wadge hierarchy of $\om$-languages 
accepted by {\it deterministic} Turing machines; 
its length is $(\om_1^{\mathrm{CK}})^\om$    
\cite{Selivanov03a,Selivanov03b}.

\hs In previous papers we have inductively constructed, using the work of Duparc 
on the Wadge hierarchy of  Borel sets \cite{Duparc01}, some ${\bf \Delta}_\om^0$ context free 
$\om$-languages in $\varepsilon_\om$ Wadge degrees, 
where $\varepsilon_\om$ is the $\om^{th}$  fixed point of the ordinal exponentiation of base $\om$,   and 
also some ${\bf \Sigma}_\om^0$-complete context free 
$\om$-languages, 
\cite{Fin01a,Fin01c,Fin03a,Fin03c}. Notice that the Wadge hierarchy of {\it non-deterministic}
context-free $\om$-languages  is not effective. 

\hs We are going to show here the very surprising following result, which extends Theorem \ref{thebor}.

\begin{The}\label{thewad}    
The Wadge hierarchy of the class 
{\bf r}-${\bf BCL}(1)_\om$, hence also of the class ${\bf CFL}_\om$, or of every 
class $\mathcal{C}$ such that 
$\mbox{ {\bf r}-}{\bf BCL}(1)_\om \subseteq  \mathcal{C} $$ \subseteq  \Sigma^1_1$, 
is the Wadge hierarchy of the class 
 $\Sigma^1_1$ of $\om$-languages accepted by Turing machines with a B\"uchi acceptance 
condition. 
\end{The}

\noi To prove this result, we  are going to consider first non self dual sets. We recall the definition of Wadge degrees 
introduced by Duparc in \cite{Duparc01} and which is a slight modification of the previous one. 

\begin{Deff}
\noi
\begin{enumerate}
\ite[(a)]  $d_w(\emptyset)=d_w(\emptyset^-)=1$
\ite[(b)]  $d_w(L)=sup \{d_w(L')+1 ~\mid ~ L' {\rm ~non~ self ~dual~ and~}
L'<_W L \} $
\nl (for either $L$ self dual or not, $L>_W \emptyset).$
\end{enumerate}
\end{Deff}

\noi  We are going now to
introduce the operation of sum of 
sets of infinite words which has as  
counterpart the ordinal
addition  over Wadge degrees.

\begin{Deff}[Wadge, see \cite{Duparc01}]
Assume that $X\subseteq Y$ are two finite alphabets,
  $Y-X$ containing at least two elements, and that
$\{X_+, X_-\}$ is a partition of $Y-X$ in two non empty sets.
 Let $L \subseteq X^{\om}$ and $L' \subseteq Y^{\om}$, then
 $$L' + L =_{df} L\cup \{ u.a.\beta  ~\mid  ~ u\in X^\star , ~(a\in X_+
~and ~\beta \in L' )~
or ~(a\in X_- ~and ~\beta \in L'^- )\}$$
\end{Deff}

\noi This operation is closely related to the {\it ordinal sum}
 as it is stated in the following:

\begin{The}[Wadge, see \cite{Duparc01}]\label{thesum}
Let $X\subseteq Y$, $Y-X$ containing at least two elements,
   $L \subseteq X^{\om}$ and $L' \subseteq Y^{\om}$ be 
non self dual  Borel sets.
Then $(L+L')$ is a non self dual Borel set and
$d_w( L'+L )= d_w( L' ) + d_w( L )$.
\end{The}

\noi A player in charge of a set $L'+L$ in a Wadge game is like a player in charge of the set $L$ but who 
can, at any step of the play,    erase  his previous play and choose to be this time in charge of  $L'$ or of $L'^-$. 
Notice that he can do this only one time during a play. We shall use this property below. 

\begin{Lem}\label{6-10}
Let $L \subseteq \Sio$ be a non self dual  Borel set such that $d_w( L )\geq \om$. Then it holds that $L \equiv_W \emptyset + L$. 
\end{Lem}

\noi Notice that in the above lemma, $\emptyset$ is viewed as the empty set over an alphabet $\Gamma$ such that 
$\Si \subseteq \Ga$ and cardinal ($\Ga - \Si$) $\geq 2$. 

\begin{proof}
Assume that  $L \subseteq \Sio$ is a non self dual  Borel set and  that $d_w( L )\geq \om$.
We know that $\emptyset$ is a non self dual Borel set and that $d_w( \emptyset ) = 1$. Thus, by Theorem \ref{thesum}, 
it holds that $d_w(  \emptyset + L ) = d_w(  \emptyset ) + d_w( L ) = 1 + d_w( L ) $. But by hypothesis $d_w( L )\geq \om$ and this implies 
that  $1 + d_w( L ) = d_w( L )$. So we have proved that $d_w(  \emptyset + L ) =  d_w( L )$. 
 \nl On the other hand $L$ is non self dual and $d_w(  \emptyset + L ) =  d_w( L )$ imply that only two cases may happen : 
either $\emptyset + L  \equiv_W  L$  or  $\emptyset + L \equiv_W L^-$. 
\nl But it is easy to see that $L \leq_W \emptyset + L$. For that purpose consider the Wadge game $W( L, \emptyset + L)$. 
Player 2 has clearly a winning strategy which consists in copying the play of Player 1 thus $L \leq_W \emptyset + L$. 
This implies that $\emptyset + L \equiv_W L^-$ cannot hold so $\emptyset + L  \equiv_W  L$. 
\ep  \end{proof}
 
\begin{Lem}\label{6-11}
Let  $L \subseteq \Sio$ be a non self dual  Borel set acccepted by a Turing machine with a B\"uchi acceptance 
condition. Then there is an $\om$-language $L' \in $  {\bf r}-${\bf BCL}(8)_\om$ such that $L \equiv_W L'$.  
\end{Lem}

\begin{proof}
It is well known that there are regular $\om$-languages of every finite Wadge degree, \cite{Staiger97,Selivanov98}. These $\om$-languages 
are Boolean combinations of open sets.  So we have only to consider the case of non self dual Borel sets of Wadge 
degrees greater than or equal to $\om$. 

\hs Let then $L \subseteq \Sio$ be a non self dual  Borel set acccepted by a Turing machine with a B\"uchi acceptance 
condition ( in particular $L$ is in the class ${\bf BCL}(2)_\om$ ) 
such that $d_w( L )\geq \om$.   
\hs Lemma \ref{theta_S} states that  there exists an integer $S \geq 1$ such 
that $\theta_S(L)$ is in the class  {\bf r}-${\bf BCL}(8)_\om$, where $E$ is a new letter not in 
$\Si$ and $\theta_S: \Sio \ra (\Sigma \cup \{E\})^\om$ is  the 
function defined, for all  $x \in \Sio$, by: 
$$\theta_S(x)=x(1).E^{S}.x(2).E^{S^2}.x(3).E^{S^3}.x(4) \ldots 
x(n).E^{S^n}.x(n+1).E^{S^{n+1}} \ldots $$

\noi We are going to prove that  $\theta_S(L) \equiv_W L$.  

\hs Firstly, it is easy to see that $L \leq_W \theta_S(L)$. In order to prove this  we can consider the Wadge game 
$W( L, \theta_S(L) )$. It is easy to see that Player 2 has a winning strategy in this game which consists in copying the play of Player 1, except that 
Player 2 adds letters $E$ in such a way that he has written the initial word 
$x(1).E^{S}.x(2).E^{S^2}.x(3).E^{S^3}.x(4) \ldots 
x(n).E^{S^n}$ while Player 1 has written the initial word $x(1).x(2).x(3).x(4) \ldots 
x(n)$. 
Notice that one can admit that a player writes a finite word at each step of the play instead of a single letter. This does not change the 
winner of a Wadge game. 

\hs To prove that $ \theta_S(L) \leq_W L $, it suffices to prove that $ \theta_S(L) \leq_W \emptyset + L $ because 
Lemma \ref{6-10} states that $ \emptyset + L \equiv_W  L$.  Consider the Wadge game $W(  \theta_S(L),  \emptyset + L )$. 
\nl Player 2 has a winning strategy in this play which consists first in copying the play of player 1 except that Player 2 skips when player 1 writes a letter $E$. 
He continues forever with this strategy if the word written by player 1 is always a prefix of some $\om$-word of $\theta_S(\Sio)$. Then after $\om$ steps 
Player 1 has written an $\om$-word $\theta_S(x)$ for some $x \in \Sio$, and Player 2 has written $x$. So in that case 
$\theta_S(x) \in  \theta_S(L)$ iff  $x \in L$ iff  $x \in \emptyset + L $.  
\nl But if at some step of the play, Player 1 ``goes out of" the closed set  $\theta_S(\Sio)$ because the word he has now 
written is not a prefix of any $\om$-word of  $\theta_S(\Sio)$, then its final word will be surely outside $\theta_S(\Sio)$ hence also outside 
$ \theta_S(L)$. 
Player 2 can now writes a letter of $\Gamma -\Si$ in such a way that he is now like a player in charge of the emptyset  and he can 
now writes an $\om$-word $u$ so that his final $\om$-word will be outside $\emptyset + L $. Thus Player 2 wins this play too. 

\hs Finally we have proved that $L \leq_W \theta_S(L)   \leq_W L $ thus it holds that $\theta_S(L) \equiv_W L$.  This ends the proof.

\ep  \end{proof}

\begin{Lem}\label{6-12}
Let  $L \subseteq \Sio$ be a non self dual  Borel set  in the class {\bf r}-${\bf BCL}(8)_\om$. Then 
 there is an $\om$-language $L' \in $  {\bf r}-${\bf BCL}(1)_\om$ such that $L \equiv_W L'$.  
\end{Lem}

\begin{proof}
As in the preceding proof we can consider only $\om$-languages of Wadge degrees greater than or equal to $\om$. 

\hs  Let then $L=L(\mathcal{A})$$\subseteq \Sio$ be a non self dual  Borel set acccepted by a real time $8$-counter B\"uchi automaton $\mathcal{A}$ 
such that $d_w( L )\geq \om$.   We have shown in the preceding section that 
$\phi_K ( h_K( L(\mathcal{A}) )$$ \cup h_K(\Si^{\om})^- )$  is in the class {\bf r}-${\bf BCL}(1)_\om$, where $h_K$ is the 
continuous mapping $h_K : \Sio \ra  (\Si \cup\{A, B, 0\})^\om$ defined  by :  for all $x\in \Sio$,
$$h_K(x)=A.0^K.x(1).B.0^{K^2}.A.0^{K^2}.x(2).B.0^{K^3}.A.0^{K^3}.x(3).B \ldots  
B.0^{K^n}.A.0^{K^n}.x(n).B \ldots  $$
\noi and  the mapping 
$\phi_K : (\Si\cup\{A, B, 0\})^\om \ra (\Si\cup\{A, B, F,  0\})^\om $ is defined by:  
for all $x\in (\Si\cup\{A, B, 0\})^\om$, 
$$\phi_K(x) = F^{K-1}.x(1).F^{K-1}.x(2).F^{K-1}.x(3) 
\ldots F^{K-1}.x(n). F^{K-1}.x(n+1).F^{K-1} \ldots$$
\noi We can now prove, by  a very similar reasoning as in the proof of the preceding Lemma \ref{6-11},  using the fact that $d_w( L )\geq \om$, that 
\begin{center}
$L \equiv_W  h_K( L(\mathcal{A}) )$$ \cup h_K(\Si^{\om})^-  \equiv_W  $ $\phi_K ( h_K( L(\mathcal{A}) )$$ \cup h_K(\Si^{\om})^- )$ 

\end{center}
\noi But $\phi_K ( h_K( L(\mathcal{A}) )$$ \cup h_K(\Si^{\om})^- )$  is in the class {\bf r}-${\bf BCL}(1)_\om$, and this ends the proof.

\ep  \end{proof}

{\it End of Proof of Theorem \ref{thewad}. }   
Let  $L \subseteq \Sio$ be a   Borel set acccepted by a Turing machine with a B\"uchi acceptance 
condition (in particular $L$ is in the class ${\bf BCL}(2)_\om$). If the Wadge degree of $L$ is finite, it is well known that it is Wadge 
equivalent to a regular $\om$-language, hence also to an  $\om$-language in the class {\bf r}-${\bf BCL}(1)_\om$. 
If  $L$ is non self dual and its Wadge degree is greater than or equal to $\om$, then we can infer from Lemmas \ref{6-11} and  \ref{6-12} that 
there is an $\om$-language  $L' \in $  {\bf r}-${\bf BCL}(1)_\om$ such that $L \equiv_W L'$. 
\nl It remains to consider the case of self dual Borel sets.  The alphabet $\Si$ being finite, a self dual Borel set $L$ is always Wadge equivalent to a Borel set 
in the form $\Si_1.L_1 \cup \Si_2.L_2$, where $(\Si_1, \Si_2)$ form a partition of $\Si$, and $L_1, L_2\subseteq \Sio$ are non self dual Borel sets such that 
$L_1 \equiv_W L_2^-$.  
Moreover $L_1$ and $L_2$ can be taken in the form $L_{(u_1)}=u_1.\Sio \cap L$ and      $L_{(u_2)}=u_2.\Sio \cap L$     for some $u_1, u_2 \in \Sis$, see
\cite{Duparc03}. 
So if  $L \subseteq \Sio$ is a self dual Borel set accepted by a  Turing machine with a B\"uchi acceptance 
condition then $L \equiv_W \Si_1.L_1 \cup \Si_2.L_2$, where $(\Si_1, \Si_2)$ form a partition of $\Si$, and 
 $L_1, L_2\subseteq \Sio$ are non self dual Borel sets accepted by a  Turing machine with a B\"uchi acceptance 
condition.  
We have already proved that there is  an $\om$-language $L'_1 \in $  {\bf r}-${\bf BCL}(1)_\om$ such that $L'_1 \equiv_W L_1$ and 
 an $\om$-language $L'_2 \in $  {\bf r}-${\bf BCL}(1)_\om$ such that $L_2'^-\equiv_W L_2$.  Thus 
$L  \equiv_W  \Si_1.L_1 \cup \Si_2.L_2  \equiv_W \Si_1.L_1' \cup \Si_2.L'_2$ and $\Si_1.L'_1 \cup \Si_2.L'_2$ is in the class 
{\bf r}-${\bf BCL}(1)_\om$. 
\ep

\begin{Rem}
We have only considered above the Wadge hierarchy of {\bf  Borel sets}. If we assume the axiom of ${\bf \Si}_1^1$-determinacy, then Theorem 
\ref{wh} can be extended by considering the class of analytic sets instead of the class of Borel sets. In fact  in that case any set which is analytic but not Borel is 
${\bf \Si}_1^1$-complete, see \cite{Kechris94}. So there is only one more Wadge degree containing  ${\bf \Si}_1^1$-complete sets. 
It was already proved in \cite{Fin03a} that 
there is a ${\bf \Si}_1^1$-complete set accepted by a B\"uchi $1$-counter automaton and it is easy to see from the proof that one can find such a 
${\bf \Si}_1^1$-complete set accepted by a B\"uchi $1$-counter {\it real-time} automaton. 

\end{Rem}

\begin{Rem}
The result given by Theorem \ref{thebor} can now be deduced from  Theorem \ref{thewad} and it can be seen as a particular case of this last result, 
because the Wadge hierarchy is a refinement of the Borel hierarchy and, 
 for each countable non null ordinal $\gamma$, ${\bf \Si}_\gamma^0$-complete sets (respectively, ${\bf \Pi}_\gamma^0$-complete sets) 
form a single equivalence class of $\equiv_W$, i.e. a single Wadge degree,  \cite{Kechris94}. 
However we have preferred to expose  the results given in this paper by considering firstly the Borel hierarchy. This way the reader who is just interested by the Borel 
hierarchy of $\om$-languages can read this part and skip Section \ref{section-wadge} about the Wadge hierarchy. 

\end{Rem}

\section{Concluding remarks}

\noi We have proved that the Borel  and the Wadge hierarchies 
of  classes {\bf r}-${\bf BCL}(1)_\om$  
and ${\bf CFL}_\om$  are also the Borel and  the Wadge hierarchies of the 
class $\Sigma^1_1$. 
The methods used in this paper are different from those used in previous papers on context free 
$\om$-languages \cite{Fin01a,Fin01c,Fin03a,Fin03c}, where we gave an inductive construction 
of some ${\bf \Delta}_\om^0$ context free $\om$-languages of a given Borel rank 
or Wadge degree, using work of Duparc 
on the Wadge hierarchy of ${\bf \Delta}_\om^0$ Borel sets, \cite{Duparc01}. However it will be 
possible to combine both methods for the effective 
construction of  $\om$-languages in the class  {\bf r}-${\bf BCL}(1)_\om$, and of   
$1$-counter B\"uchi automata accepting them,  
of a given Wadge degree among the $\varepsilon_\om$ degrees 
obtained in \cite{Fin01c} for ${\bf \Delta}_\om^0$ context free $\om$-languages. 
\nl   Finally we mention that in another paper, using  the results of this paper and applying  similar methods to the 
study of topological properties of infinitary rational relations,  we prove 
that their Wadge and Borel hierarchies are equal to the corresponding hierarchies 
of the classes 
{\bf r}-${\bf BCL}(1)_\om$,  ${\bf CFL}_\om$  or $\Sigma^1_1$, \cite{Fin04b}. 

\section*{Acknowledgements}
Thanks to the anonymous referee for useful comments
on a preliminary version of this paper.

\end{document}